\documentclass[a4paper,12pt]{article}
\usepackage{amsmath}
\usepackage{a4wide}
\usepackage{amssymb}
\usepackage{epsfig}
\usepackage{amsfonts}
\usepackage{color}
\usepackage{tikz}

\newtheorem{lemma}{Lemma}
\newtheorem{theorem}{Theorem}
\newtheorem{definition}{Definition}
\newtheorem{proposition}{Proposition}

\newtheorem{corollary}{Corollary}
\newtheorem{algorithm}{Algorithm}

\newtheorem{example}{Example}
\newenvironment{proof}{\par\noindent\underline{Proof}: }{\hfill$\square$\par\vskip10pt}

\newcommand{\Cross}{\mathbin{\tikz [x=1.4ex,y=1.4ex,line width=.2ex] \draw (0,0) -- (1,1) (0,1) -- (1,0);}}%
\begin{document}

\title{The structure of strategy-proof rules} 
\author{Jorge Alcalde--Unzu\thanks{Department of Economics and INARBE, Universidad P\'ublica de Navarra, Campus Arrosadia, 31006 Pamplona, Spain. \texttt{Email:\,jorge.alcalde@unavarra.es}. Financial support from the Spanish Ministry of Economy and Competitiveness, through the project PID2021-127119NB-I00, is gratefully acknowledged.}\, and Marc Vorsatz\thanks{Corresponding author. Departamento de An\'alisis Econ\'omico, Universidad Nacional de Educaci\'on a Distancia (UNED), Paseo Senda del Rey 11, 28040 Madrid, Spain. \texttt{Email:\,mvorsatz@cee.uned.es}. Financial support from  the Spanish Ministry of Economy and Competitiveness, through the project PID2021-122919NB-I00, is gratefully acknowledged.} \vspace{0.3cm}}
\thispagestyle{empty}
\maketitle
\begin{abstract}\vspace{0.1cm}
\renewcommand{\baselinestretch}{1.2}\normalsize
\noindent 
We establish that all strategy-proof social choice rules in strict preference domains follow necessarily a two-step procedure. In the first step, agents are asked to reveal some specific information about their preferences. Afterwards, a subrule that is dictatorial or strategy-proof of range 2 must be applied, and the selected subrule may differ depending on the answers of the first step. %Formally, this structure follows from two findings: ($i$) the Gibbard-Satterthwaite impossibility extends to {\it non-conditional} preference domains (those in which each agent has a set of fixed binary relations), and ($ii$) any preference domain that is not non-conditional can be converted into a non-conditional one whenever the agent reveals some information about her true preference.
As a consequence, the strategy-proof rules that have been identified in the literature for some domains can be reinterpreted in terms of our procedure and, more importantly, this procedure serves as a guide for determining the structure of the strategy-proof rules in domains that have not been explored yet.
\vspace{0.3cm}
\vspace{0.3cm}
\par
\noindent \textit{JEL-Numbers:} D70, D71.
\end{abstract}

\newpage
\renewcommand{\baselinestretch}{1.5}\normalsize
\section{Introduction}

Consider any problem of collective choice in which a group of agents has to choose an alternative from a finite set. Since, usually, agents' preferences are private information and the chosen alternative depends on the declared preferences, some agents may have incentives to misrepresent their preferences. The social choice rules that are immune to this type of strategic manipulation are said to be {\it strategy-proof}: specifically, they satisfy the condition that independently of the other agents' preferences, it is always optimal for an agent to declare her true preference.

\medskip

\noindent The difficulty of constructing strategy-proof social choice rules is reflected by the seminal Gibbard-Satterthwaite impossibility theorem \cite{Gib,Sat}, which establishes that if preferences are unrestricted, all strategy-proof rules are either dictatorial or of range 2. One interpretation of this result is that the universal domain contains a vast amount of strategies so that there is always room for profitable manipulations. As a consequence, the literature has been aiming at identifying restricted domains in which the impossibility disappears. However, thus far there are no results common to all restricted domains and, therefore, characterizing the strategy-proof rules in a restricted domain can be a complex task.

\medskip

\noindent The main objective of this paper is to uncover a general procedure that permits the construction of all strategy-proof rules for the cartesian product of any combination of strict preference domains. This structure is obtained by combining three facts: ($i$) we introduce a new classification of strict preference domains into conditional and non-conditional preference domains; ($ii$) we show that the Gibbard-Satterthwaite impossibility still holds when the preference domain of each agent is non-conditional; and ($iii$) we identify a procedure through which any conditional preference domain can be further restricted to become non-conditional. These three facts together allow us to conclude that any strategy-proof rule consists of two steps. First, each agent with a conditional preference domain is asked to reveal some specific information about her preference in such a way that we can apply the procedure in ($iii$) and her preference domain becomes non-conditional. This first step guarantees that we arrive at a situation in which our extension of the Gibbard-Satterthwaite impossibility in ($ii$) applies. The second step thus consists of a subrule that is either dictatorial or of range 2. Note that the applied subrule in this second step can depend on the information provided in the first step. Although all strategy-proof rules necessarily comply with this two-step procedure, there are rules with this structure that are manipulable. However, if the second-step subrules are either dictatorial or strategy-proof of range 2, the only possible manipulations are in the first step. Consequently, to obtain a complete characterization of all strategy-proof rules, the {\it only} task is to analyze the combinations of second-step subrules that avoid manipulations in the first step.

\medskip

\noindent To illustrate our main findings with an example, let there be three alternatives $x$, $y$, and $z$ so that the universal strict preference domain is ${\cal R} = \{xyz, xzy, yxz, yzx, zxy, zyx\}$.\footnote{Here, we represent each preference ranking by an ordered list. For example, $xyz$ is the preference ranking according to which $x$ is preferred to $y$ and $y$ is preferred to $z$.} Consider the following two restricted preference domains.
\begin{itemize}
\item[(a)] ${\cal R}'$ includes all preferences (and only those) that satisfy the statement ``$x$ is preferred to $z$''. One can think of this particular case as a situation in which it is publicly known that the agent prefers $x$ to $z$, yet there is complete uncertainty about the remaining preference structure, that is, ${\cal R}' = \{xyz, yxz, xzy\}$. Restricted preference domains that are obtained from the universal preference domain by fixing a set of binary comparisons are referred to as \emph{non-conditional preference domains}. In the case of ${\cal R}'$, the only fixed comparison is $(x, z)$.

\item[(b)] ${\cal R}''$ includes all preferences (and only those) that satisfy the statement ``if $x$ is preferred to $y$, then $y$ is preferred to $z$''. The preferences $xzy$ and $zxy$ are thus excluded from the preference domain, that is, ${\cal R}'' = \{xyz, yxz, yzx, zyx\}$. This preference domain is obtained for instance if alternatives represent locations on the real line, with $x < y < z$, and the agent has single-peaked preferences over the locations. If a set of antecedents needs to hold for some pairwise comparison to be fixed, then we speak of a \emph{conditional preference domain}. In the case of ${\cal R}''$, the conclusion ``$y$ is preferred to $z$'' only follows if the antecedent ``$x$ is preferred to $y$'' is true. 
\end{itemize}

\medskip

\noindent To see how one can construct all strategy-proof rules in any setting, suppose that there are only two agents $i$ and $j$ such that the preference domain of agent $i$ is equal to ${\cal R}'$, while the preference domain of agent $j$ is ${\cal R}''$. Observe that our extension of the impossibility result does not apply directly because the preference domain of agent $j$ is conditional. However, the crucial observation is that if it was additionally known whether $j$ prefers $x$ to $y$ or $y$ to $x$ (\emph{i.e.}, if it was additionally known whether the actual preference ranking of $j$ satisfies the antecedent of the conditional restriction), then her preference domain would be non-conditional as well.
If $j$ actually prefers $x$ to $y$, then ${\cal R}''$ can be further restricted to $\{xyz\}$, which is the non-conditional preference domain associated with the set of fixed comparisons $\{(x, y), (y, z), (x, z)\}$. 
And if $j$ actually prefers $y$ to $x$, her preference domain is $\{yxz, yzx, zyx\}$, which is the non-conditional preference domain associated with the set of fixed comparisons $\{(y, x)\}$.
So, once the actual preference of $j$ between $x$ and $y$ is known, we arrive at a situation in which the preference domains of both agents are non-conditional. Thus, it follows from our impossibility that any strategy-proof social choice rule that operates after the preference of agent $j$ on the pair $\{x,y\}$ has been truthfully revealed is either of range 2 or dictatorial.
Therefore, any strategy-proof social choice rule $f$ in this example should initially ask agent $j$ about how she effectively orders $x$ and $y$. Depending on the answer, the social choice rule $f$ uses then a particular subrule $f_x$ (if $j$ says she prefers $x$ to $y$) or $f_y$ (if $j$ says she prefers $y$ to $x$) such that each of the subrules is strategy-proof of range 2 or dictatorial. Since the set of strategy-proof subrules of range 2 and dictatorial subrules is fairly easy to describe, the only task is to check which combinations of $f_x$ and $f_y$ from this set can be applied without giving agent $j$ the possibility to manipulate the outcome by misrepresenting her preference between $x$ and $y$ in the first step. This strategy is general and can be applied to any situation, independently of the number of agents and alternatives involved and the particular strict preference domain of each agent.

\medskip

\noindent Our paper contributes to the existing literature on strategy-proofness in three ways. While several authors have generalized the Gibbard-Satterthwaite impossibility by focusing on common preference domains, our impossibility also applies to personalized preference domains (see the concluding section for a discussion). Second, influential studies like Moulin \cite{Mou}, Sprumont \cite{Spr}, and Barber\`a, Sonnenschein and Zhou \cite{BarSonZho} have characterized meaningful strategy-proof rules on specific domains (see also Barber\`a \cite{Bar} for a survey). These rules can be reinterpreted in terms of our two-step procedure.  Finally, and most importantly, the two-step procedure serves as a guide for determining the structure of the strategy-proof rules on domains that have not been explored yet.

\medskip

\noindent The remainder is organized as follows. Section 2 introduces notation and presents the impossibility result. We also apply the impossibility result to problems in which all agents have non-conditional preference domains. In Section 3, we derive the structure of all strategy-proof rules if the preference domains of some agents are conditional. We conclude in Section 4. The proof of the impossibility result is in the Appendix.

\section{Impossibility result}

\noindent Let $N$ and $X$ be the finite sets of agents and alternatives, respectively. The preference of each agent $i \in N$ over $X$ is described by a complete, antisymmetric, and transitive binary relation $R_i$. Let $P_i$ be the strict preference relation induced by $R_i$. The universal preference domain ${\cal R}$ includes all complete, antisymmetric, and transitive preference relations over $X$. We will work with (possibly) restricted preference domains that can differ across agents; {\it i.e.}, we denote by ${\cal R}_i \subseteq {\cal R}$ the preference domain of agent $i \in N$, and we refer to $\Cross_{i \in N} {\cal R}_i = {\cal D}$ as the domain. A preference profile $R = (R_i)_{i \in N} \in {\cal D}$ is a list of agents' preferences.

\medskip

\noindent Given a domain ${\cal D}$, a \emph{social choice rule} is a function $f: {\cal D} \rightarrow X$ that selects for each preference profile $R \in {\cal D}$ an alternative $f(R) \in X$. We denote the range of $f$ by $r(f)$. 
Let $R_{-S}$ be the subprofile obtained by eliminating the set of agents $S$ from profile $R$. We write $R_{-i}$ instead of $R_{-\{i\}}$.
We say that $f$ is {\it manipulable by agent} $i \in N$ if there is a preference profile $R \in {\cal D}$ and an alternative preference $R'_i \in {\cal R}_i$ such that $f(R'_i, R_{-i}) \, P_i \, f(R)$. Then, $f$ is {\it strategy-proof} (SP) if it is not manipulable by any agent. We say that $f$ is a {\it dictatorship of agent} $i \in N$ if for all $R \in {\cal D}$ and all $x \in r(f), f(R) \, R_i \, x$. 
Then, $f$ is {\it non--dictatorial} (ND) if there is no agent $i \in N$ such that $f$ is a dictatorship of agent $i$. 
If preferences are unrestricted, Gibbard \cite{Gib} and Satterthwaite \cite{Sat} have shown the impossibility of combining SP and ND whenever $|r(f)| \neq 2$.\footnote{The impossibility is sometimes presented with the condition $|r(f)| \geq 2$. Since $|r(f)| = 1$ is incompatible with ND (if $|r(f)| = 1$, all agents are dictators), we use the weaker requirement $|r(f)| \neq 2$ instead.  Other versions of the impossibility result require that $|X| >2$ and that $f$ is unanimous (\emph{i.e.}, if the top alternative of all agents' rankings is the same, then this alternative must be selected by $f$). The combination of these two conditions is again stronger than the requirement $|r(f)| \neq 2$.}

\begin{proposition}
\label{old}
(Gibbard \cite{Gib} and Satterthwaite \cite{Sat}) If ${\cal R}_i = {\cal R}$ for all $i \in N$, there is no social choice rule $f$ with $|r(f)| \neq 2$ that is SP and ND.
\end{proposition}

\noindent Our impossibility generalizes Proposition \ref{old}. Let $X^* = \{(x, y) \in X^2 \, | \, x \neq y\}$ be the set of all ordered pairs of distinct alternatives. Given any preference domain ${\cal R}_i$, we denote the set of ordered pairs that are fixed for all preferences belonging to ${\cal R}_i$ by $S({\cal R}_i)$; \emph{i.e.}, $S({\cal R}_i) = \{(x, y) \in X^* \; | \; x \, P_i \, y \mbox{ for all } R_i \in {\cal R}_i\}$. Then, a preference domain ${\cal R}_i$ for agent $i$ is {\em non-conditional} whenever the following condition holds: $R_i \in {\cal R}_i$ if and only if $x \, P_i \, y$ for all $(x, y) \in S({\cal R}_i)$. Otherwise, ${\cal R}_i$ is {\em conditional}. We also define $S^{-1}({\cal R}_i) = \{\{x, y\} \in X^2 \; | \; (x, y),(y,x) \not\in S({\cal R}_i) \mbox { and } x \neq y\}$.
$S^{-1}({\cal R}_i) $ denotes the set of pairs that can be ordered freely in ${\cal R}_i$. Two extreme cases of non-conditional preference domains are the universal preference domain (since $S({\cal R})$ is empty) and the preference domains ${\cal R}_i$ that contains only one admissible preference ranking (where $S^{-1}({\cal R}_i)$ is empty). Our result establishes that the Gibbard-Satterthwaite impossibility extends to non-conditional preference domains, independently of whether the domain is common or personalized.\footnote{A domain is common if ${\cal R}_i = {\cal R}_j$ for all $i, j \in N$. Otherwise, it is said to be personalized.}

\begin{theorem}
\label{new}
Suppose that for all $i \in N$, ${\cal R}_i$ is a non-conditional preference domain. Then, there is no rule $f$ with $|r(f)| \neq 2$ that is SP and ND.
\end{theorem}

\noindent Theorem \ref{new} can be directly applied to problems whose natural characteristics imply that agents' preferences are non-conditionally restricted. For example, consider a set of agents $N$ that wants to assign an indivisible good to one of its members. Standard economic theory assumes that each agent prefers the object to be assigned to herself rather than to any other agent, which implies that each agent $i \in N$ has a non-conditional preference domain ${\cal R}_i$ with fixed pairs $S({\cal R}_i) = \{(i, j) \mbox{ for all } j \in N \setminus \{i\}\}$. It follows then from our result that the Gibbard-Satterthwaite impossibility is maintained in this domain and, thus, all SP rules are of range 2 or are dictatorial. This application is related to Holzman and Moulin \cite{HM} who assume that each agent nominates some candidate different from herself. They focus on the property of impartiality (\emph{i.e.,} the outcome should be independent of the message of the winning candidate), for which they show some possibility results. Since impartiality is weaker than strategy-proofness if agents reported strict preferences instead of nominating a particular candidate (this is because impartiality only eliminates manipulations that make oneself the receptor of the indivisible good and does not consider manipulations further down in the ranking), it follows from our impossibility result that the strengthening of impartiality to strategy-proofness on this domain severely impacts the possibility of constructing meaningful social choice rules. In a second application, which has some features in common with Amor\'os \cite{A1, A2}, a set of jurors with publicly known a priori biases has to choose a winning candidate. In case of a juror $i \in N$ with a bias in favor of the candidates from the set $A_i \subset X$ over the candidates from the set $B_i \subseteq X \setminus A_i$, the preference domain of this juror can be restricted to the non-conditional domain ${\cal R}_i$ with fixed pairs $S({\cal R}_i) = \{(x, y) \mbox{ for all } x \in A_i \mbox{ and } y \in B_i\}$. Theorem \ref{new} shows that the Gibbard-Satterthwaite impossibility cannot be avoided in these instances.

\section{SP rules in conditional domains}

\noindent Theorem \ref{new} can only be directly applied to non-conditional domains. In this section, we show that Theorem \ref{new} also provides a route for the construction of all SP rules in the remaining domains; {\it i.e.}, when the preference domain of at least one agent is conditional. We divide our argument into three parts. First, while Section 2 defines a conditional preference domain as any preference domain that is not non-conditional, we now define a conditional preference domain by means of a set of specific restrictions that are jointly applied to the universal preference domain. 
This allows us to show, secondly, how any conditional preference domain can be further restricted to become non-conditional. And, finally, we detail how this procedure of restricting a conditional preference domain to a non-conditional one, together with Theorem \ref{new}, identifies the structure of the SP rules in the cartesian product of any combination of strict preference domains.

\subsection*{Preference domain restrictions}

\noindent We introduce two ways in which a preference domain can be restricted. 

\begin{definition}
\label{non-cond}
Given ${\cal R}_i, {\cal R}'_i$, with ${\cal R}'_i \subset {\cal R}_i$, and an ordered pair $(x, y)$,
\begin{itemize}
\item ${\cal R}'_i$ is a $(x, y)$ non-conditional restriction of ${\cal R}_i$ whenever for any $R_i \in {\cal R}_i$, 
$$R_i \not\in {\cal R}'_i \mbox{ if and only if } y \, P_i \, x.$$

\item ${\cal R}'_i$ is a conditional restriction of ${\cal R}_i$ whenever there is a set of ordered pairs of alternatives $\{(x_k, y_k)\}_{k = 1}^K$ such that for any $R_i \in {\cal R}_i$,
$$R_i \not\in {\cal R}'_i \mbox{ if and only if } x_k \, P_i \, y_k \mbox{ for all } k \in \{1, \ldots, K\} \mbox{ and } y \, P_i \, x.$$
\end{itemize}

\end{definition}

\noindent We highlight that a non-conditional restriction eliminates all rankings from a preference domain that do not order a pair of alternatives in a pre-described way, while a conditional restriction consists of an antecedent (the ordered pairs $\{(x_k, y_k)\}_{k = 1}^K$) and a conclusion (the ordered pair $(x, y)$) such that a ranking is eliminated if it satisfies the antecedent but does not meet the conclusion. Algorithm 1 below shows that any preference domain can be obtained from the universal preference domain via the iterative application of these two types of restrictions. To be precise, if the considered preference domain is non-conditional, it is only necessary to apply a chain of non-conditional restrictions, meanwhile if it is conditional, it is necessary to apply, apart from maybe some non-conditional restrictions, a set of conditional restrictions.

\begin{algorithm} 
Consider any preference domain ${\cal R}_i \subseteq {\cal R} \equiv {\cal R}_i^0$.

\medskip%\\*[-10pt]

\noindent \underline{Step $t = 1, 2, \ldots, \frac{|X|(|X|-1)}{2}$}:

\medskip
 
\noindent Take any unordered pair of distinct alternatives $\{x^t,y^t\}$ that has not been considered in a previous step. If $(x^t, y^t) \in S({\cal R}_i)$ or if $(y^t, x^t) \in S({\cal R}_i)$, then go to step $t.a$. If $\{x^t,y^t\} \in S^{-1}({\cal R}_i)$, then go to step $t.b$.

\begin{itemize}
\item [($t.a$)] If $(x^t,y^t) \in S({\cal R}_i)$, ${\cal R}_i^t$ is set equal to the $(x^t, y^t)$ non-conditional restriction of ${\cal R}_i^{t-1}$. 

If $(y^t, x^t) \in S({\cal R}_i)$, ${\cal R}_i^t$ is set equal to the $(y^t, x^t)$ non-conditional restriction of ${\cal R}_i^{t-1}$.

\item [($t.b$)] Set ${\cal R}_i^t = {\cal R}_i^{t-1}$.
\end{itemize}

\noindent If ${\cal R}_i = {\cal R}_i^t$, then the algorithm ends and ${\cal R}_i$ is declared to be a  non-conditional preference domain. If ${\cal R}_i \subset {\cal R}_i^t$, then proceed with step $t+1$.\\*[-10pt]

\noindent \underline{Step $t > \frac{|X|(|X|-1)}{2}$}:\\
\noindent Take any $R_i \in {\cal R}_i^{t-1} \setminus {\cal R}_i$ and any unordered pair of distinct alternatives $\{x^t,y^t\} \in S^{-1}({\cal R}_i)$ such that $y^t \, P_i \, x^t$. Construct a minimal set of ordered pairs $\{(x_k, y_k)\}_{k = 1}^K$ such that for all $R'_i \in {\cal R}_i^{t-1}$ with $x_k \, P'_i \, y_k$ for all $k \in \{1, \ldots, K\}$, $$R'_i \in {\cal R}_i \Leftrightarrow x^t \, P'_i \, y^t.$$
Apply the conditional restriction with antecedent $\{(x_k, y_k)\}_{k = 1}^K$ and conclusion $(x^t, y^t)$ to the preference domain ${\cal R}_i^{t-1}$ and denote the resulting preference domain by ${\cal R}_i^t$. If ${\cal R}_i = {\cal R}_i^t$, then the algorithm ends and ${\cal R}_i$ is declared to be a  conditional preference domain. If ${\cal R}_i \subset {\cal R}_i^t$, then proceed with step $t+1$.
\end{algorithm}

\noindent Since the set of alternatives $X$ is finite, Algorithm 1 always terminates and therefore classifies each strict preference domain as conditional or non-conditional. Even more importantly, Algorithm 1 describes a conditional preference domain as a set of conditional and non-conditional restrictions. Note that the set of conditional restrictions may not be unique because a change in the selection of the preference $R_i \in {\cal R}_i^{t-1} \setminus {\cal R}_i$ and the unordered pair $\{x^t,y^t\} \in S^{-1}({\cal R}_i)$ in a step $t > \frac{|X|(|X|-1)}{2}$ may lead to different conditional restrictions that define the same preference domain. For example, consider the single-peaked preference domain over the triple $\{x, y, z\}$ with order $x < y < z$;\footnote{A preference $R_i$ is \emph{single-peaked} if there exists an alternative $s \in X$ (called the \emph{peak}) such that for all $t, u \in X$, [$s \geq t > u$ or $u > t \geq s$] imply $t \, P_i \, u$.} {\it i.e.}, the set of rankings $\{xyz, yxz, yzx, zyx\}$. One way to obtain this preference domain is through the conditional restriction with antecedent $(x, y)$ and conclusion $(y, z)$. Another way is through the conditional restriction with antecedent $(z, y)$ and conclusion $(y, x)$. This multiplicity is an advantage because each way provides a different but independent approach to the construction of all SP rules.

\medskip

\noindent Similarly, there are occasions in which different sets of non-conditional restrictions have the same implications. For example, since preferences are transitive, the sets of non-conditional restrictions $\{(x, y), (y, z), (x, z)\}$ and $\{(x, y), (y, z)\}$ describe the same non-conditional preference domain. Moreover, each non-conditional restriction could be substituted by a pair of conditional restrictions with inverse antecedents: {\it e.g.}, the non-conditional restriction $(x, y)$ can be replaced by the union of two conditional restrictions, one with antecedent $(z, w)$, other with antecedent $(w, z)$ and both with conclusion $(x, y)$. However, the construction of the SP rules is simpler if fewer conditional restrictions are imposed, and this is the reason why Algorithm 1 prioritizes the non-conditional restrictions in its first steps.

\medskip

\noindent To complete this subsection, we define any preference domain in terms of Algorithm 1. In particular, ${\cal R}_i$ is defined through a mapping $f_{{\cal R}_i} : 2^{X^*} \cup \emptyset \rightarrow 2^{X^*} \cup \emptyset$ such that for all $(x, y) \in X^*$,

\begin{itemize}
\item $(x, y) \in f_{{\cal R}_i}(\emptyset)$ if and only if the $(x, y)$ non-conditional restriction has been applied to classify ${\cal R}_i$ through Algorithm 1.

\item for any $\{(x_k, y_k)\}_{k=1}^K \in 2^{X^*}$, we have that $(x, y) \in f_{{\cal R}_i}(\{(x_k, y_k)\}_{k=1}^K)$ if and only if the conditional restriction with antecedent $\{(x_k, y_k)\}_{k = 1}^K$ and conclusion $(x, y)$ has been applied to classify ${\cal R}_i$ through Algorithm 1.
\end{itemize}

\noindent The mapping $f_{{\cal R}_i}$ defines the preference domain ${\cal R}_i$, yet note that this identifying mapping is not unique because, as we have argued before, different sets of conditional restrictions may be applied. For each such mapping $f_{{\cal R}_i}$, let $C(f_{{\cal R}_i}) = \{(x, y) \in X^* \, | \, \exists A \in 2^{X^*} \mbox{ such that } (x, y) \in A \mbox{ and } f_{{\cal R}_i}(A) \neq \emptyset\}$ be the set of all ordered pairs that belong to the antecedents of the conditional restrictions that define ${\cal R}_i$ according to $f_{{\cal R}_i}$. Consequently, if $C(f_{{\cal R}_i}) = \emptyset$, then ${\cal R}_i$ is non-conditional. Otherwise, ${\cal R}_i$ is conditional.

\subsection*{Restricting conditional preference domains}

\noindent Next, we show how a conditional preference domain can be further restricted to become non-conditional. Consider any conditional preference domain ${\cal R}_i$ identified by $f_{{\cal R}_i}$. If it was known whether the antecedents of the conditional restrictions that define ${\cal R}_i$ are satisfied ({\it i.e.}, if the actual preference of agent $i$ over the pairs belonging to $C(f_{{\cal R}_i})$ were known), then it would be possible to further restrict the preference domain of agent $i$ from ${\cal R}_i$ to a non-conditional preference domain. 
In particular, if it was known that the true preference of the agent $R_i \in {\cal R}_i$ satisfies exactly $B \subseteq C(f_{{\cal R}_i})$ of the ordered pairs of the antecedents, then it would be possible to further restrict ${\cal R}_i$ to ${\cal R}_i^B$, where $f_{{\cal R}_i^B}(A) = \emptyset$ for all $A \in 2^{X^*}$ and $f_{{\cal R}_i^B}(\emptyset) = f_{{\cal R}_i}(\emptyset) \cup B \cup (C(f_{{\cal R}_i}) \setminus B)^{-1} \cup \{f_{{\cal R}_i}(D)\}_{D \subseteq B}$ with $(C({\cal R}_i) \setminus B)^{-1} = \{(x, y) \in X^* \, | \, (y, x) \in (C({\cal R}_i) \setminus B)\}$. Note that $C(f_{{\cal R}_i^B}) = \emptyset$ and, thus, ${\cal R}_i^B$ is a non-conditional domain. As an illustration, we apply this procedure in two examples.

\begin{example}
\emph{Suppose that Algorithm 1 has determined that  ${\cal R}_i$ does not contain any non-conditional restriction and exactly two conditional restrictions, both with antecedent $(x, y)$ and conclusions $(z, w)$ and $(z, v)$, respectively. That is, $f_{{\cal R}_i}(\{(x, y)\}) = \{(z, w), (z, v)\}$ and for all $A \in 2^{X^*} \setminus \{(x, y)\}$, $f_{{\cal R}_i}(A) = f_{{\cal R}_i}(\emptyset) = \emptyset$. Thus, $C(f_{{\cal R}_i}) = \{(x, y)\}$ and ${\cal R}_i$ is conditional. Then, if it was known whether the true preference $R_i$ satisfies the antecedent $(x, y)$, it would be possible to further restrict ${\cal R}_i$ to become non-conditional.}

\begin{itemize}
\item[($i$)] \emph{If $R_i$ satisfies the antecedent $(x,y)$, then $z$ must be ranked above $w$ and $v$. Thus, ${\cal R}_i^{(x, y)} \subset {\cal R}_i$ only has the non-conditional restrictions $\{(x, y), (z, w), (z, v)\}$. That is, $f_{{\cal R}_i^{(x, y)}}(\emptyset) = \{(x, y), (z, w), (z, v)\}$ and for all $A \in 2^{X^*}$, $f_{{\cal R}_i^{(x, y)}}(A) = \emptyset$. Since $C(f_{{\cal R}_i^{(x, y)}}) = \emptyset$, ${\cal R}_i^{(x, y)}$ is non-conditional.}
 
\item[($ii$)] \emph{If $R_i$ does not satisfy the antecedent $(x,y)$, then $z$ may or may not be ranked above $w$ and $v$. Thus, $R_i^{\emptyset} \subset {\cal R}_i$  only has the non-conditional restriction $\{(y, x)\}$. That is, $f_{{\cal R}_i^{\emptyset}}(\emptyset) = \{(y, x)\}$ and for all $A \in 2^{X^*}$, $f_{{\cal R}_i^{\emptyset}}(A) = \emptyset$. Since $C(f_{{\cal R}_i^{\emptyset}}) = \emptyset$, ${\cal R}_i^{\emptyset}$ is also non-conditional.} $\hfill \square$
\end{itemize}
\end{example}

\begin{example}
\emph{Let ${\cal R}_j$ be the preference domain of all single-peaked preferences over the set $\{v, w, x, y, z\}$ with order $v < w < x < y < z$. The function $f_{{\cal R}_j}$, where $f_{{\cal R}_j}(\{(v, w)\}) = \{(w, x)\}$, $f_{{\cal R}_j}(\{(w, x)\}) = \{(x, y)\}$, $f_{{\cal R}_j}(\{(x, y)\}) = \{(y, z)\}$, and $f_{{\cal R}_j}(A) = f_{{\cal R}_j}(\emptyset) = \emptyset$ for all $A \in 2^{X^*} \setminus [\{(v, w)\} \cup \{(w, x)\} \cup \{(x, y)\}]$ identifies this preference domain and it can be obtained from Algorithm 1. Thus, $C(f_{{\cal R}_j}) = \{(v, w), (w, x), (x, y)\}$. In order to learn which combination of these antecedents is satisfied by the true preference $R_j$ of the agent, it is sufficient to know whether the peak of $R_j$ is at $v$, $w$, $x$, or to the right of $x$.}

\begin{itemize}
\item[($i$)] \emph{If $R_j$ has the peak at $v$, then all three antecedents are satisfied. Thus, the preference domain ${\cal R}_j^B \subset {\cal R}_j$, where $B = \{(v, w), (w, x), (x, y)\}$, only has the non-conditional restrictions $\{(v, w), (w, x), (x, y), (y, z)\}$. Hence, $f_{{\cal R}_j^B}(\emptyset) = \{(v, w), (w, x), (x, y), (y, z)\}$ and for all $A \in 2^{X^*}$, $f_{{\cal R}_j^B}(A) = \emptyset$. Since $C(f_{{\cal R}_j^B}) = \emptyset$, ${\cal R}_j^B$ is non-conditional.}

\item[($ii$)] \emph{If $R_j$ has the peak at $w$, then only the antecedents $(w, x)$ and $(x, y)$ are satisfied. Thus, ${\cal R}_j^B \subset {\cal R}_j$, where $B = \{(w, x), (x, y)\}$, only has the non-conditional restrictions $\{(w, v), (w, x), (x, y), (y, z)\}$. Consequently, $f_{{\cal R}_j^B}(\emptyset) = \{(w, v), (w, x), (x, y), (y, z)\}$ and for all $A \in 2^{X^*}$, $f_{{\cal R}_j^B}(A) = \emptyset$. Since $C(f_{{\cal R}_j^B}) = \emptyset$, ${\cal R}_j^B$ is also non-conditional.}

\item[($iii$)] \emph{If $R_j$ has the peak at $x$, then only the antecedent $(x, y)$ is satisfied. Thus, ${\cal R}_j^{(x, y)} \subset {\cal R}_j$ only has the non-conditional restrictions $\{(w, v), (x, w), (x, y),$ $(y, z)\}$. Consequently, $f_{{\cal R}_j^{(x, y)}}(\emptyset) = \{(w, v), (x, w), (x, y), (y, z)\}$ and for all $A \in 2^{X^*}$, $f_{{\cal R}_j^{(x, y)}}(A) = \emptyset$. Since $C({\cal R}_j^{(x, y)}) = \emptyset$, ${\cal R}_j^{(x, y)}$ is non-conditional.}

\item[($iv$)] \emph{If the peak of $R_j$ is to the right of $x$, then no antecedent is satisfied. Thus, the preference domain ${\cal R}_j^{\emptyset} \subset {\cal R}_j$ only has the non-conditional restrictions $\{(w, v), (x, w), (y, x)\}$. Consequently, $f_{{\cal R}_j^{\emptyset}}(\emptyset) = \{(w, v), (x, w), (y, x)\}$ and for all $A \in 2^{X^*}$, $f_{{\cal R}_j^{\emptyset}}(A) = \emptyset$. Since $C(f_{{\cal R}_j^{\emptyset}}) = \emptyset$, ${\cal R}_j^{\emptyset}$ is also non-conditional.} $\hfill \square$
\end{itemize}
\end{example}

\subsection*{The structure of SP rules}

\noindent The preceding subsection establishes that a conditional preference domain can be further restricted to become non-conditional if some particular information about the agent's preference is obtained. Furthermore, Section 2 shows that the Gibbard-Satterthwaite impossibility also applies to the cartesian product of any combination of non-conditional preference domains. As a consequence of these two facts, any SP social choice rule on the cartesian product of any combination of strict preference domains is characterized by the following two-step procedure:

\begin{enumerate}
\item Information is extracted from each agent $i$ in order to deduce which antecedents of the conditional restrictions that define her preference domain ${\cal R}_i$ according to Algorithm 1 are satisfied by the actual preference. This information is always obtained if the individual is asked about how she ranks all ordered pairs of $C(f_{{\cal R}_i})$ for some $f_{{\cal R}_i}$. As a result of this first step, the preference domain of each agent is non-conditional.

\item For each combination of non-conditional preference domains that can arise from the first step, one must apply by Theorem \ref{new} a subrule that is either dictatorial or that is strategy-proof of range 2.
\end{enumerate}

\noindent Although this two-step decomposition is necessary for a social choice rule to be SP, it is not a complete characterization of the set of all SP rules: the second-step subrules depend on the information provided in the first step and, therefore, one must construct combinations of subrules in the second step that are consistent with truthful preference revelation in the first step.
As we will show next with the help of examples, in spite of not being a complete characterization, our procedure, apart from being applicable to all domains, not only substantially reduces the number of rules that are potentially SP, it also limits the manipulations that should be checked to obtain a characterization.
We elaborate on Examples 1 and 2 assuming that $N=\{1,2\}$ and $X = \{v, w, x, y, z\}$.

\medskip

\noindent {\bf Example 1 (continuation)}: Suppose that for all $i \in \{1, 2\}$, $f_{{\cal R}_i}((x, y)) = \{(z, w), (z, v)\}$ and $f_{{\cal R}_i}(A) = f_{{\cal R}_i}(\emptyset) = \emptyset$ for any $A \in 2^{X^*} \setminus \{(x, y)\}$. The first step of any SP rule asks each agent whether her actual preference satisfies the antecedent $(x, y)$. Afterwards, a subrule, which may depend on the response profile of the first step, that is either dictatorial or strategy-proof of range 2 must be applied. The calculations in the Appendix show that this two-step procedure reduces the number of potentially SP rules from a total of $5^{6400}$ to only $37 \cdot 46^2 \cdot 59$. Apart from this substantial reduction, our procedure reduces the possible manipulations to misrepresentations of the true preference only on the pair $\{x, y\}$. Thus, to obtain a complete characterization, it is sufficient to check which of these rules do not give incentives to misrepresent the true preferences on that pair. $\hfill \square$

\medskip

\noindent {\bf Example 2 (continuation)}: Suppose that for all $j \in \{1, 2\}$, $f_{{\cal R}_j}(\{(v, w)\}) = \{(w, x)\}$, $f_{{\cal R}_j}(\{(w, x)\}) = \{(x, y)\}$, $f_{{\cal R}_j}(\{(x, y)\}) = \{(y, z)\}$, and for all $A \in 2^{X^*} \setminus [\{(v, w)\} \cup \{(w, x)\} \cup \{(x, y)\}]$, $f_{{\cal R}_j}(A) = f_{{\cal R}_j}(\emptyset) = \emptyset$. The first step of any SP rule asks each agent whether her peak is at $v$, $w$, $x$ or to the right of $x$.  Afterwards, a subrule, which may depend on the response profile of the first step, that is either dictatorial or strategy-proof of range 2 must be applied. The calculations in the Appendix show that this two-step procedure reduces the number of potentially SP rules from a total of $5^{256}$ to only $21^2 \cdot 17^3 \cdot 16^2 \cdot 14^2 \cdot 9^4 \cdot 8^2 \cdot 5$. Moreover, the only manipulations that might occur are misrepresentations of preferences in the first step and, therefore, to obtain a complete characterization, it is sufficient to check which of these rules do not give incentives to misrepresent the true preferences in this first step. $\hfill \square$

\medskip

\noindent Observe that the domains in Examples 1 and 2 are common. However, our procedure is also applicable if the domain is personalized. An interesting example along this line is Alcalde-Unzu and Vorsatz \cite{AUV} who study the situation when each agent either has single-peaked or single-dipped preferences\footnote{A preference $R_i$ is \emph{single-dipped} if there exists an alternative $s \in X$ (called the \emph{dip}) such that for all $t, u \in X$, [$s \geq t > u$ or $u > t \geq s$] imply $u \, P_i \, t$.}  and the type of the preference is private information, but the location of her peak/dip is public information and can vary across agents. This domain emerges when society has to choose the location of a public facility which provokes different opinions: agents that like/dislike the facility have single-peaked/dipped preferences, and the peak/dip of the preference of the agent coincides naturally with her own location, which is known to the decision maker. These preference domains are conditional, but if the types of preferences of the agents were known, then the preference domains would become non-conditional. It follows thus from our characterization strategy that any SP rule first asks each agent about her type of preferences (single-peaked or single-dipped). Then, for each response profile of the first step, a dictatorial subrule or a strategy-proof subrule of range 2 has to be applied in the second step. So, it is only necessary to find the combinations of subrules that induce agents to reveal their types of preferences truthfully in the first step. Alcalde-Unzu and Vorsatz \cite{AUV} follows exactly this approach to obtain the characterization of all SP rules in that domain.

\section{Conclusion}

\noindent We have shown in this paper that all strategy-proof social choice rules on the cartesian product of any combination of strict preference domains necessarily can be described with a two-step procedure. In the first step, agents are asked to reveal some specific information about their preferences. Afterwards, a subrule that is dictatorial or strategy-proof of range 2 must be applied. To be more concrete, agents with conditional preference domains are asked first how they effectively order the pairs of alternatives that belong to the antecedents of the conditional restrictions that define their preference domains. As a consequence, their conditional preference domains become non-conditional. In the second step, our extension of the Gibbard-Satterthwaite impossibility for non-conditional preference domains operates and this is the reason why the only strategy-proof subrules are dictatorships or of range 2. 

\medskip

\noindent Our findings imply that any strategy-proof rule that has been identified in the literature can be reinterpreted in terms of this two-step procedure and, more importantly, this two-step procedure serves as a guide for determining the structure of the strategy-proof rules on domains that have not been explored yet. The literature has also considered some stronger properties than strategy-proofness like \emph{group strategy-proofness} (see, Barber\`a, Berga and Moreno \cite{BBM, BBM2}) or \emph{obviously strategy-proofness} (see, Li \cite{Li}). Our procedure can also be applied to characterize the set of rules that satisfy these stronger properties. For example, the group strategy-proof rules necessarily comply with our two-step procedure and for a complete characterization it is only necessary to analyze which combinations of subrules of the second step are immune to not only individual but also to group manipulations in the first step.

\medskip

\noindent Our main formal instrument is Theorem \ref{new}, which is a generalization of the Gibbard-Satterthwaite theorem. Other extensions of this seminal result are due to Kalai and Muller \cite{KM}, Blair and Muller \cite{BM}, Aswal, Chatterji and Sen \cite{ACS}, Sato \cite{Sato}, Reffgen \cite{Ref}, and Pramanik \cite{Pra}. The main difference between Theorem \ref{new} and this literature is that these other generalizations focus on some particular classes of common domains, meanwhile Theorem \ref{new} also applies to personalized domains. And this difference turns out to be crucial. Indeed, some of the existing generalizations (for instance, Aswal, Chatterji and Sen \cite{ACS}) imply that the Gibbard-Satterthwaite impossibility also holds when all agents have the \emph{same} non-conditional preference domain. However, our strategy to construct the structure of all SP rules in any domain needs the impossibility to hold true also when the non-conditional preference domains differ across agents. The reason is that once information about the antecedents of the conditional restrictions is obtained in the first step, we could arrive at a situation in which agents have different non-conditional preference domains. And this could occur even if all agents share originally a specific conditional preference domain, simply because the agents might differ in their responses to the antecedents of the conditional restrictions. 

\medskip

\noindent Finally, note that for some domains the two-step procedure is not necessarily the most compact way of describing its SP rules: for instance, although all the generalized median voter rules in the single-peaked preference domain can be defined with the structure laid out in Example 2, this is most likely not the most elegant approach. Thus, there are domains for which, after obtaining all SP rules by means of the two-step procedure, some additional work must be carried out to obtain an elegant description of the family. However, this is the price to pay for introducing a general structure that is valid in all domains. Moreover, there are domains, such as the one analyzed in Alcalde-Unzu and Vorsatz \cite{AUV}, for which our two-step procedure is the natural characterization approach. 

%\medskip

%\noindent Second, there is also an interesting interpretation of Theorem \ref{new} in terms of the roots of the original Gibbard-Satterthwaite impossibility. In the universal domain, neither does the social planner have any information about the actual preferences of the agents nor does the set of alternatives have any pre-defined structure ({\it i.e.}, like ordering alternatives on the real line). To see which of these two characteristics is decisive for the impossibility, observe that contexts in which the social planner knows how some agents actually compare some alternatives lead to non-conditional preference domains, meanwhile if some structure is imposed on the set of alternatives {\color{blue} (\emph{e.g.}, single-peaked preferences), this might result in a conditional preference domain}. Consequently, Theorem \ref{new} shows that the Gibbard-Satterthwaite impossibility is not due to a lack of information about the agents' true preferences, but rather a problem of a missing structure on the set of alternatives. 

\newpage

\section*{Appendix}

\noindent We first introduce some notation. 
We define $O_i(R_{-i}, {\cal R}_i) = \{x \in X \; | \; \exists R_i \in {\cal R}_i \mbox{ such that }$ $f(R_i, R_{-i}) = x\}$ as the option set of agent $i$ with a preference domain ${\cal R}_i$ when the other agents have preferences $R_{-i}$. 
We also define ${\cal D}_{-i} = \Cross_{j \in N \setminus \{i\}} {\cal R}_j$. 
Let $R_i|_S$ be the preference that is obtained when $R_i$ is restricted to the set of alternatives $S \subset X$. 
Then, $\max\{R_i|_{O_i(R_{-i}, {\cal R}_i)}\}$ is the most preferred alternative of agent $i$ from her option set $O_i(R_{-i}, {\cal R}_i)$ when she has preferences $R_i$. 
Before proving Theorem \ref{new}, we are going to present three lemmas that will help us in the proof.

\begin{lemma}
\label{maximal}
If $f$ is SP on ${\cal D}$, then for all $R \in {\cal D}$ and all $i \in N$, $f(R) = \max\{R_i|_{O_i(R_{-i}, {\cal R}_i)}\}$. \end{lemma}

\begin{proof}
\noindent Suppose otherwise, that is, there is an agent $i \in N$ and a preference profile $R \in {\cal D}$ such that $f(R) \neq \max\{R_i|_{O_i(R_{-i}, {\cal R}_i)}\}$. 
Since $\max\{R_i|_{O_i(R_{-i}, {\cal R}_i)}\} \in O_i(R_{-i}, {\cal R}_i)$, there is a preference $R'_i \in {\cal R}_i$ such that $f(R'_i, R_{-i}) = \max\{R_i|_{O_i(R_{-i}, {\cal R}_i)}\}$. 
Given that $f(R) \in O_i(R_{-i}, {\cal R}_i)$ by definition, we have that $\max\{R_i|_{O_i(R_{-i}, {\cal R}_i)}\} \, P_i \, f(R)$. 
Thus, agent $i$ manipulates $f$ at $R$ via $R'_i$. 
This contradicts that $f$ is SP on $\cal D$.
\end{proof}

\noindent It follows from Lemma \ref{maximal} that an agent $j$ is a dictator at a SP rule $f$ on domain ${\cal D}$ if for all $R_{-j} \in {\cal D}_{-j}$, $O_j(R_{-j}, {\cal R}_j) = r(f)$. The second lemma shows that if $f$ is SP on some domain ${\cal D}$, then it remains to be SP on any subdomain ${\cal D}'$ of ${\cal D}$; that is, whenever ${\cal R}'_i \subseteq {\cal R}_i$ for all $i \in N$. We denote the restriction of $f$ from ${\cal D}$ to ${\cal D}'$ by $f'$.

\begin{lemma}
\label{fact}
If $f$ is SP on ${\cal D}$ and ${\cal D}'$ is a subdomain of ${\cal D}$, then $f'$ is SP on ${\cal D}'$.
\end{lemma}

\begin{proof}
Suppose by contradiction that $f'$ is not SP on ${\cal D}'$. 
Then, there is a profile $R \in {\cal D}'$, an agent $i \in N$, and a preference $R'_i \in {\cal R}'_i$ such that $f'(R'_i, R_{-i}) \, P_i \, f'(R)$. 
Since ${\cal D}'$ is a subdomain of ${\cal D}$ by assumption, it follows from $R \in {\cal D}'$ that $R \in {\cal D}$, and from $(R'_i, R_{-i}) \in {\cal D}'$ that $(R'_i, R_{-i}) \in {\cal D}$. 
Then, since $f'$ is the restriction of $f$ from ${\cal D}$ to ${\cal D}'$, $f(R'_i, R_{-i}) = f'(R'_i, R_{-i}) \, P_i \, f'(R) = f(R)$. 
Agent $i$ thus manipulates $f$ at $R$ via $R'_i$. 
This contradicts that $f$ is SP on $\cal D$.
\end{proof}

\noindent The third lemma shows that all pairs of alternatives that belong to the option set of an agent are free pairs in her preference domain.

\begin{lemma}
\label{free} 
Suppose that $f$ is SP on ${\cal D}$. 
Then, for all $i \in N$, all $R_{-i} \in {\cal D}_{-i}$, and all $x,y \in O_i(R_{-i}, {\cal R}_i)$, $\{x, y\} \in S^{-1}({\cal R}_i)$.
\end{lemma}

\begin{proof}
\noindent Suppose otherwise, that is, there is an agent $i \in N$, a preference profile for the other agents $R_{-i} \in {\cal D}_{-i}$, and two alternatives $x, y \in O_i(R_{-i}, {\cal R}_i)$ such that $\{x, y\} \not\in S^{-1}({\cal R}_i)$. 
Then, by definition of $S^{-1}({\cal R}_i)$, $(x, y) \in S({\cal R}_i)$ or $(y, x)\in S({\cal R}_i)$. 
Suppose without loss of generality that $(x, y) \in S({\cal R}_i)$. 
Since  $x, y \in O_i(R_{-i}, {\cal R}_i)$ by assumption, there are two preferences $R_i, R'_i \in {\cal R}_i$ such that $f(R_i, R_{-i}) = x$ and $f(R'_i, R_{-i}) = y$. 
Agent $i$ thus manipulates $f$ at $(R'_i, R_{-i})$ via $R_i$. 
This contradicts that $f$ is SP on $\cal D$.
\end{proof}

\subsection*{Proof of Theorem \ref{new}}

\noindent The proof is induction--based. 
If $S^{-1}({\cal R}_i)= \emptyset$ for all $i \in N$, then the size of the range of any rule $f$ is $|r(f)|=1$. 
Therefore, there is no rule $f$ with $|r(f)| \neq 2$ that is SP and ND. 
This is the base case of the induction.
Next, suppose that the impossibility holds for some non-conditional domain ${\cal D}' = {\cal R}'_i \times {\cal D}'_{-i}$ and consider a larger non-conditional domain ${\cal D} = {\cal R}_i \times {\cal D}'_{-i}$ that satisfies the following condition: for some alternatives $x,y \in X$ such that $(x, y) \in S({\cal R}'_i)$ and such that there is no $z \in X$ with $(x, z), (z, y) \in S({\cal R}'_i)$, $S({\cal R}_i) = S({\cal R}'_i) \setminus (x, y)$. 
The induction step requires us to show that the impossibility also holds for ${\cal D}$.\footnote{Note that by means of this induction argument, it is possible to arrive at any non-conditional domain ${\cal D}$.}

\bigskip

\noindent Suppose by contradiction that there is a rule $f$ on $\cal D$ with $|r(f)| \neq 2$ that is both SP and ND. 
By Lemma \ref{fact}, $f'$ is SP on ${\cal D}'$. 
Since the impossibility holds by assumption on ${\cal D}'$, it must be the case that either $|r(f')| = 2$ or $f'$ is dictatorial. We divide the proof into three parts. First, we establish two essential lemmas that additionally show that $|r(f')| \neq 1$. Then, we show that $f'$ cannot be dictatorial with $|r(f')| \neq 1$. Finally, we prove that $f'$ cannot be ND with $|r(f')| = 2$.

\bigskip

\noindent {\it PART 1: Two lemmas}

\medskip

\noindent We first introduce some notation.
Consider any pair of alternatives $w, z \in X$ and define, associated with this pair, a permutation $\pi_{w, z}$ of $X$ such that $\pi_{w, z}(w) = z$, $\pi_{w, z}(z) = w$, and $\pi_{w, z}(v) = v$ for all $v \not\in \{w, z\}$. Then, for each preference relation $R_j$ of some agent $j \in N$, we construct the preference $R_j^{\pi_{w,z}}$ such that $u \, R_j \, v \Leftrightarrow \pi_{w, z}(u) \, R_j^{\pi_{w,z}} \, \pi_{w, z}(v)$ for all $u, v \in X$. Thus, $R_j^{\pi_{w,z}}$ is obtained from $R_j$ by only switching alternatives $w$ and $z$. We denote $R_j^{\pi_{w,z}}$ by $R_j(w,z)$ whenever $w \, P_j^{\pi_{w,z}}  z$ and by $R_j(z,w)$ whenever $z \, P_j^{\pi_{w,z}} w$. For a given preference $R_j$ such that $w \, P_j \, z$, let $K(w, z, R_j) = \{v \in X \; | \; w \, R_j \, v \, R_j \, z\}$ be the set of alternatives that are at least as good as $z$ but not strictly better than $w$ at $R_j$ (that is, $w$ and $z$ are included in this set) and let $\hat{K}(w, z, R_j) = \{v \in X \; | \; w \, P_j \, v \, P_j \, z\}$ be the set of alternatives that are strictly ranked between $w$ and $z$ at $R_j$. Finally, $T({\cal R}_j, w, z) = \{v \in X \; | \; \{w, v\}, \{z, v\} \in S^{-1}({\cal R}_j)\}$ denotes the set of alternatives that agent $j \in N$ can rank freely with respect to both $w$ and $z$ under the preference domain ${\cal R}_j$.

\medskip

\noindent Consider any preference $R_i \in {\cal R}_i \setminus {\cal R}'_i$. Since $S({\cal R}_i)=S({\cal R}'_i) \setminus (x,y)$ by the induction hypothesis, $R_i \in {\cal R}_i \setminus {\cal R}'_i$ implies that $y \, P_i \, x$. 
Lemma \ref{relation} then shows that (a) the preference $R_i(x,y)$, which is obtained from $R_i$ by only permuting $x$ and $y$, must belong to ${\cal R}'_i$, and (b) given any alternative $z$ that is strictly ranked between $y$ and $x$ at $R_i$, both $\{x, z\}$ and $\{y, z\}$ are free pairs under the preference domain ${\cal R}'_i$. 

\begin{lemma}
\label{relation}
If $R_i \in {\cal R}_i \setminus {\cal R}'_i$, then $\hat{K}(y, x, R_i) \subseteq T({\cal R}'_i, x, y)$ and $R_i(x,y) \in {\cal R}'_i$.
\end{lemma}

\begin{proof} We first show that for each $R_i \in {\cal R}_i \setminus {\cal R}'_i$, $\hat{K}(y, x, R_i) \subseteq T({\cal R}'_i, x, y)$. 
Suppose by contradiction that there is an alternative $z \in \hat{K}(y, x, R_i) \setminus T({\cal R}'_i, x, y)$. 
Then, by definition, either $\{x, z\}$ or $\{y, z\}$ does not belong to $S^{-1}({\cal R}'_i)$. 
Suppose that $\{x, z\} \not\in S^{-1}({\cal R}'_i)$ (the proof of the other case is similar). 
There are two possibilities. 
First, if $(z, x) \in S({\cal R}'_i)$, then $(z, y) \in S({\cal R}'_i)$ by transitivity and, thus, $(z, y) \in S({\cal R}_i)$. This contradicts that $z \in \hat{K}(y, x, R_i)$. 
Second, if $(x, z) \in S({\cal R}'_i)$, then $(x, z) \in S({\cal R}_i)$, which again contradicts that $z \in \hat{K}(y, x, R_i)$. 
It follows finally from $\hat{K}(y, x, R_i) \subseteq T({\cal R}'_i, x, y)$ that $R_i(x,y) \in {\cal R}'_i$. \end{proof}

\noindent Note that for all subprofiles $R_{-i} \in {\cal D}_{-i}$, $O_i(R_{-i}, {\cal R}'_i) \subseteq O_i(R_{-i}, {\cal R}_i)$. 
The next crucial lemma analyzes the relation between these option sets when they differ.

\begin{lemma}
\label{range0}
For each $R_{-i} \in {\cal D}_{-i}$ such that $O_i(R_{-i}, {\cal R}'_i) \neq O_i(R_{-i}, {\cal R}_i)$, $\{x, y\} \cap O_i(R_{-i}, {\cal R}'_i) = x$ and $O_i(R_{-i}, {\cal R}_i) = O_i(R_{-i}, {\cal R}'_i) \cup \{y\}$.
\end{lemma}

\begin{proof}
Let $R_{-i} \in {\cal D}_{-i}$ be such that $O_i(R_{-i}, {\cal R}'_i) \neq O_i(R_{-i}, {\cal R}_i)$ and let $z \in X$ be such that $z \in O_i(R_{-i}, {\cal R}_i) \setminus O_i(R_{-i}, {\cal R}'_i)$.
Since $z \in O_i(R_{-i}, {\cal R}_i) \setminus O_i(R_{-i}, {\cal R}'_i)$, there is a preference $R_i \in {\cal R}_i \setminus {\cal R}'_i$ such that $f(R_i,R_{-i})=z$.
Now, consider the preference $R_i(x,y)$.
By Lemma \ref{relation}, $R_i(x,y) \in {\cal R}'_i$. 
Then, by construction, $f(R_i(x,y), R_{-i}) \neq z$. 
Denote $f(R_i(x,y), R_{-i})=w$.  
We show that $z=y$ and $w=x$ by reaching a contradiction in the remaining cases. We divide the proof into three steps.

\begin{enumerate}
\item [1.] \emph{Suppose that $\{w, z\} \cap \{x, y\} = \emptyset$ or $\{w, z\} \not\subseteq K(y, x, R_i)$.}\smallskip

If $\{w, z\} \cap \{x, y\} = \emptyset$ or $\{w, z\} \not\subseteq K(y, x, R_i)$, then $w \, P_i \, z \Leftrightarrow w \, P_i(x,y) \, z$. 
Agent $i$ thus manipulates $f$ at $(R_i, R_{-i})$ via $R_i(x,y)$ whenever $w \, P_i \, z$ and at $(R_i(x,y), R_{-i})$ via $R_i$ whenever $z \, P_i \, w$ (and, therefore, $z \, P_i(x,y) \, w$). 
This contradicts that $f$ is SP.

\item [2.] \emph{Suppose that $[w = y$ and $z \in K(y,x,R_i) \setminus \{y\}]$ or $[z = x$ and $w \in K(y,x,R_i) \setminus \{x\}]$.}\smallskip

Suppose that $w = y$ and $z \in K(y,x,R_i) \setminus \{y\}$ (the proof of the other case is similar). Given that $z \in K(y,x, R_i) \setminus \{y\}$, $z \in K(x, y, R_i(x,y)) \setminus \{y\}$. Then, $z \, P_i(x, y) \, y = w$. Agent $i$ thus manipulates $f$ at $(R_i(x,y), R_{-i})$ via $R_i$. 
This contradicts that $f$ is SP.

\item[3.] \emph{Suppose that $[z = y$ and $w \in \hat{K}(y, x, R_i)]$ or $[w = x$ and $z \in \hat{K}(y, x, R_i)]$.}\smallskip

Suppose that $z = y$ and $w \in \hat{K}(y, x, R_i)$ (the proof of the other case is similar). Consider the preference $\hat{R}_i$ such that (a) for all $u, v \in X \setminus \{y\}$, $u \, R_i(x,y) \, v \Leftrightarrow u \, \hat{R}_i \, v$ and (b) for all $s \in X \setminus \{y\}, y \, \hat{P}_i \, s \Leftrightarrow w \, R_i(x,y) \, s$. That is, $\hat{R}_i$ is obtained from $R_i(x,y)$ by locating alternative $y$ just above $w$. 
By Lemma \ref{relation}, $\{y, v\} \in S^{-1}({\cal R}'_i)$ for all $v \in \hat{K}(y, x, R_i)$. Then, since $R_i(x,y) \in {\cal R}'_i$, we have that $\hat{R}_i \in {\cal R}'_i$. 
Since $z = y$ and $z \not \in O_i(R_{-i}, {\cal R}'_i)$ by assumption, we know that $f(\hat{R}_i, R_{-i}) \neq y$.
If $y \, \hat{P}_i \, f(\hat{R}_i, R_{-i})$, then agent $i$ manipulates $f$ at $(\hat{R}_i, R_{-i})$ via $R_i$. If, however, $f(\hat{R}_i, R_{-i}) \, \hat{P}_i \, y$, then $f(\hat{R}_i, R_{-i}) \, P_i(x, y) \, w$ and agent $i$ manipulates $f$ at $(R_i(x,y),R_{-i})$ via $\hat{R}_i$. 
In both cases, there is a contradiction with SP.
\end{enumerate}
\noindent This concludes the proof of the lemma.
\end{proof}

\noindent Since $r(f) = \bigcup_{R_{-i} \in {\cal D}_{-i}} O_i(R_{-i}, {\cal R}_i)$ and $r(f') = \bigcup_{R_{-i} \in {\cal D}_{-i}} O_i(R_{-i}, {\cal R}'_i)$, we can deduce the following corollary from Lemma \ref{range0}.

\begin{corollary}
\label{range}
If $r(f) \neq r(f')$, then $x \in r(f')$ and $r(f) = r(f') \cup \{y\}$. 
\end{corollary}

\noindent If $|r(f')| = 1$, Corollary \ref{range} implies that $|r(f)| \leq 2$. Since it follows from $|r(f)| = 1$ that $f$ is dictatorial (any agent acts as a dictator) and since $|r(f)| = 2$ has also been excluded by assumption, we cannot have that $|r(f')| = 1$.
So, suppose from now on that $|r(f')| \geq 2$ and, thus, $|r(f)| > 2$.

\bigskip

\noindent {\it PART 2: Suppose that $f'$ is dictatorial and that $|r(f')| \geq 2$.}

\medskip

\noindent  Since $|r(f')| \geq 2$, $f'$ cannot have more than one dictator. We divide the proof depending on the identity of the dictator.

\medskip

\noindent {\it PART 2.1: Suppose that the dictator in $f'$ is agent $i$.} \medskip

\noindent Since agent $i$ is the dictator in $f'$, it is the case that for all $R_{-i} \in {\cal D}'_{-i}$, $O_i(R_{-i}, {\cal R}'_i) = r(f')$. 
If $r(f) = r(f')$, then $O_i(R_{-i}, {\cal R}_i) = r(f)$ for all $R_{-i} \in {\cal D}'_{-i}$, which, by Lemma \ref{maximal}, implies that agent $i$ is a dictator in $f$. 
This contradicts that $f$ is ND.
Suppose therefore that $r(f) \neq r(f')$. By Corollary \ref{range}, $r(f) = r(f') \cup \{y\}$. By Lemma \ref{range0}, the set ${\cal D}'_{-i}$ can be partitioned into two subsets: the set $U$, which contains all subprofiles $R_{-i} \in {\cal D}'_{-i}$ such that $O_i(R_{-i}, {\cal R}_i) = r(f')$, and the set $V$, which contain the remaining subprofiles $\bar{R}_{-i} \in {\cal D}'_{-i}$ for which $O_i(\bar{R}_{-i}, {\cal R}_i) = r(f)$. 
Observe that $V \neq \emptyset$. 
If $U = \emptyset$, we have that $O_i(R_{-i}, {\cal R}_i) = r(f)$ for all $R_{-i} \in {\cal D}_{-i}$, which, by Lemma \ref{maximal}, implies that $f$ is a dictatorship of agent $i$, which again contradicts that $f$ is ND. 
Suppose therefore that $U \neq \emptyset$.

\smallskip

\noindent Consider a pair of subprofiles $R_{-i} \in U$ and $\bar{R}_{-i} \in V$. Then, starting at $R_{-i}$, construct a sequence of subprofiles of ${\cal D}'_{-i}$ changing one-by-one the preferences of all agents $j \neq i$ from $R_j$ to $\bar{R}_j$ so that the sequence ends at $\bar{R}_{-i}$. 
Then, there is a set $S \subset N \setminus \{i\}$ and an agent $k \not \in S \cup \{i\}$ such that $(\bar{R}_S, R_{-(S \cup \{i\})}) \in U$ and $(\bar{R}_{S \cup \{k\}}, R_{-(S \cup \{i, k\})}) \in V$. 
That is, the option set for agent $i$ is $r(f')$ if the preferences of all agents belonging to $S$ have been changed, but it is $r(f)$ if the preferences of all agents in $S \cup \{k\}$ are switched. 
Given that $|r(f')| \geq 2$ and $r(f') \cap \{x,y\} =\{x\}$, there is an alternative $z \neq y$ such that $z \in r(f') \setminus \{x\}$. 
Since $O_i(\bar{R}_{-i}, {\cal R}_i) = r(f)$, we have by Lemma \ref{free} that $\{v, w\} \in S^{-1}({\cal R}_i)$ for all $v, w \in r(f)$. Consider then two preferences $R_i, R'_i \in {\cal R}_i$ for agent $i$ such that $y \, P_i \, z \, P_i \, v$ for all $v \in r(f') \setminus \{z\}$ and $y \, P'_i \, x \, P'_i \, w$ for all $w \in r(f') \setminus \{x\}$.

\begin{enumerate}
\item [1.] \emph{We show that $\{(y, x), (z, y), (z, x)\} \cap S({\cal R}_k) \neq \emptyset$ and $\{(x, y), (y, z), (x, z)\} \cap S({\cal R}_k) \neq \emptyset$.}\smallskip

\noindent We only show that $\{(y, x), (z, y), (z, x)\} \cap S({\cal R}_k) \neq \emptyset$, because the proof of the other statement is similar. Suppose by contradiction that $\{(y, x), (z, y), (z, x)\} \cap S({\cal R}_k) = \emptyset$. Then, there is a preference $R'_k \in {\cal R}_k$ such that $x \, P'_k \, y \, P'_k \, z$.

\noindent If $(\bar{R}_S, R_{-(S \cup \{i, k\})}, R'_k) \in U$, then $O_i((\bar{R}_S, R_{-(S \cup \{i, k\})}, R'_k), {\cal R}_i) = r(f')$. Since $z \, P_i \, v$ for all $v \in r(f') \setminus \{z\}$, we have by Lemma \ref{maximal} that $f(\bar{R}_S, R_{-(S \cup \{k\})}, R'_k) = z$. 
Observe that it follows from $(\bar{R}_{S \cup \{k\}}, R_{-(S \cup \{i, k\})}) \in V$ that $O_i((\bar{R}_{S \cup \{k\}}, R_{-(S \cup \{i, k\})}), {\cal R}_i) = r(f') \cup \{y\}$. 
Given that $y \, P_i \, v$ for all $v \in r(f')$, we have by Lemma \ref{maximal} that $f(\bar{R}_{S \cup \{k\}}, R_{-(S \cup \{k\})}) = y$. Then, agent $k$ manipulates $f$ at $(\bar{R}_S, R_{-(S \cup \{k\})}, R'_k)$ via $\bar{R}_k$.

\noindent If $(\bar{R}_S, R_{-(S \cup \{i, k\})}, R'_k) \in V$, then $O_i((\bar{R}_S, R_{-(S \cup \{i, k\})}, R'_k), {\cal R}_i) = r(f') \cup \{y\}$. 
Since $y \, P'_i \, w$ for all $w \in r(f')$, we have by Lemma \ref{maximal} that $f(\bar{R}_S, R_{-(S \cup \{i, k\})}, R'_{\{i, k\}}) = y$. It also follows from $(\bar{R}_S, R_{-(S \cup \{i\})}) \in U$ that $O_i((\bar{R}_S, R_{-(S \cup \{i\})}), {\cal R}_i) = r(f')$. 
Given that $x \, P'_i \, w$ for all $w \in r(f') \setminus \{x\}$, we have by Lemma \ref{maximal} that $f(\bar{R}_S, R_{-(S \cup \{i\})}, R'_i) = x$. 
Then, agent $k$ manipulates $f$ at $(\bar{R}_S, R_{-(S \cup \{i, k\})}, R'_{\{i, k\}})$ via $R_k$.

\noindent In any case, there is a contradiction with SP.

\item [2.] \emph{We show that $\{(x, y), (z, y), (y, x), (y, z)\} \cap S({\cal R}_k) = \emptyset$.}\smallskip

\noindent Suppose by contradiction that $(x, y) \in S({\cal R}_k)$; the proofs for the other pairs follow a similar reasoning and are thus omitted.
Remember that $O_i((\bar{R}_S, R_{-(S \cup \{i\})}), {\cal R}_i) = r(f')$ and $O_i((\bar{R}_{S \cup \{k\}}, R_{-(S \cup \{i, k\})}), {\cal R}_i) = r(f') \cup \{y\}$ by construction. 
Given that $y \, P'_i \, x \, P'_i \, w$ for all $w \in r(f') \setminus \{x\}$, we obtain by Lemma \ref{maximal} that $f(\bar{R}_S, R_{-(S \cup \{i\})}, R'_i) = x$ and $f(\bar{R}_{S \cup \{k\}}, R_{-(S \cup \{i, k\})}, R'_i) = y$. 
Since $(x, y) \in S({\cal R}_k)$, $x \, \bar{P}_k \, y$. 
Agent $k$ thus manipulates $f$ at $(\bar{R}_{S \cup \{k\}}, R_{-(S \cup \{i, k\})}, R'_i)$ via $R_k$.  
This contradicts that $f$ is SP.
\end{enumerate}

\noindent It follows from 1.\ and 2.\ that $\{(x, z), (z, x)\} \subseteq S({\cal R}_k)$. This contradicts that $S({\cal R}_k)$ is antisymmetric.

\bigskip

\noindent {\it PART 2.2: Suppose that the dictator in $f'$ is agent $j \neq i$.} 

\medskip 

\noindent Since agent $j \neq i$ is the dictator in $f'$ by assumption, we can conclude that for all $R_{-i} \in {\cal D}'_{-i}$, $|O_i(R_{-i}, {\cal R}'_i)| = 1$. 
If $O_i(R_{-i}, {\cal R}'_i) = O_i(R_{-i}, {\cal R}_i)$ for all $R_{-i} \in {\cal D}'_{-i}$, then agent $j$ is also a dictator in $f$, which contradicts that $f$ is ND. 
Therefore, there is a subprofile $R_{-i} \in {\cal D}'_{-i}$ such that $O_i(R_{-i}, {\cal R}'_i) \neq O_i(R_{-i}, {\cal R}_i)$.  
Then, by Lemma \ref{range0}, $O_i(R_{-i}, {\cal R}'_i) = x$ and $O_i(R_{-i}, {\cal R}_i) = \{x, y\}$. 
Since $O_i(R_{-i}, {\cal R}'_i) = x$ and $f'$ is a dictatorship of agent $j$, $x \, P_j \, z$ for all $z \in r(f') \setminus \{x\}$. Let $R_i \in {\cal R}'_i$ and $R'_i \in {\cal R}_i \setminus {\cal R}'_i$. 
Since, by construction, $x \, P_i \, y$ and $y \, P'_i \, x$, we have by Lemma \ref{maximal} that $f(R) = x$ and $f(R'_i, R_{-i}) = y$. 
We distinguish two cases.

\begin{enumerate}

\item [1.] \emph{Suppose that $r(f) = r(f')$.}\smallskip

Since $f'$ is a dictatorship of agent $j$, we have that for all $R_{-j} \in {\cal D}'_{-j}$, $O_j(R_{-j}, {\cal R}_j) = r(f') = r(f)$. 
It follows then from Lemma \ref{free} and from $|r(f)| > 2$ that for all $z \in r(f) \setminus \{x, y\}$, $\{x, y\}, \{x, z\}, \{y, z\} \in S^{-1}({\cal R}_j)$. 
Thus, there is a preference $\hat{R}_j \in {\cal R}_j$ which is such that for all $z \in r(f) \setminus \{x, y\}$, $x \, \hat{P}_j \, z \, \hat{P}_j \, y$. 
Since agent $j$ is a dictator in $f'$ and since $(\hat{R}_j, R_{-j}) \in {\cal D}'$, $f(\hat{R}_j, R_{-j}) = x$ and $O_i((\hat{R}_j,R_{-\{i, j\}}), {\cal R}'_i) = x$.  
By Lemma \ref{range0}, $O_i((\hat{R}_j,R_{-\{i, j\}}), {\cal R}_i) \in \{\{x\}, \{x, y\}\}$ and, thus, $f(R'_i,\hat{R}_j,R_{-\{i, j\}}) \in \{x, y\}$. 
If $f(R'_i, \hat{R}_j, R_{-\{i, j\}}) = y$, consider any preference $R'_j \in {\cal R}_j$ such that for some $z \in r(f) \setminus \{x,y\}$ and  all $v \in r(f) \setminus \{z\}$, $z \, P'_j \, v$. 
Since agent $j$ is a dictator in $f'$ and since $(R'_j, R_{-j}) \in {\cal D}'$, $f(R'_j, R_{-j}) = z$ and $O_i((R'_j, R_{-\{i,j\}}), {\cal R}'_i) = z$.
Then, by Lemma \ref{range0}, $O_i((R'_j, R_{-\{i,j\}}), {\cal R}_i)=z$. This implies that $f(R'_{\{i, j\}}, R_{-\{i,j\}}) = z$. 
Agent $j$ then manipulates $f$ at $(R'_i, \hat{R}_j, R_{-\{i, j\}})$ via $R'_j$. 
If $f(R'_i, \hat{R}_j, R_{-\{i, j\}}) = x$, then agent $j$ manipulates $f$ at $(R'_i, R_{-i})$, where the outcome is $y$, via $\hat{R}_j$ (observe that $x \, P_j \, z$ for all $z \in r(f') \setminus \{x\}$ implies that $x \, P_j \, y$). 
In any case, there is a contradiction with SP.

\item [2.] \emph{Suppose that $r(f) \neq r(f')$.}\smallskip

By Corollary \ref{range}, $r(f) = r(f') \cup \{y\}$. Since $f'$ is a dictatorship of agent $j$, $O_j(R_{-j}, {\cal R}_j) = r(f')$ for all $R_{-j} \in {\cal D}'_{-j}$. Then, by Lemma \ref{free}, $\{v, w\} \in S^{-1}({\cal R}_j)$ for all $v, w \in r(f')$. We divide the analysis into two cases.

\begin{enumerate}

\item [a.] \emph{Suppose that there is an alternative $z \in r(f')$ such that $(y, z) \in S({\cal R}_j)$.}\smallskip

Consider any $w \in r(f') \setminus \{z\}$. We already know that $\{v, w\} \in S^{-1}({\cal R}_j)$ for all $v \in r(f') \setminus \{w\}$. Moreover, we have that $(v, y) \not\in S({\cal R}_j)$ for all $v \in r(f')$ because otherwise we would obtain by transitivity that $(v, z) \in S({\cal R}_j)$, which contradicts that $\{v, z\} \in S^{-1}({\cal R}_j)$. Then, there is a preference $R'_j \in{\cal R}_j$ such that for all $v \in r(f') \setminus \{w\}$, $y \, P'_j \, w \, P'_j \, v$.

\smallskip

To complete this part of the proof, note that $O_i((R'_j, R_{-\{i, j\}}), {\cal R}'_i) = w$ because $j$ is a dictator at ${\cal D}'$. Then, by Lemma \ref{range0}, $O_i((R'_j, R_{-\{i, j\}}), {\cal R}_i) = w$. 
Thus, $f(R'_{\{i, j\}}, R_{-\{i, j\}}) = w$. 
Agent $j$ thus manipulates $f$ at $(R'_{\{i, j\}}, R_{-\{i, j\}})$ via $R_j$ to obtain $y$. 
This contradicts that $f$ is SP.

\smallskip

\item[b.] \emph{Suppose that for all alternatives $z \in r(f')$, $(y, z) \not\in S({\cal R}_j)$.}\smallskip 

Consider any $z \in r(f') \setminus \{x\}$. Since $(y, v) \not\in S({\cal R}_j)$ and $\{u, v\} \in S^{-1}({\cal R}_j)$ for all $u, v \in r(f')$, there is a preference $\hat{R}_j \in {\cal R}_j$ such that $x \, \hat{P}_j \, z \, \hat{P}_j \, w \, \hat{P}_j \, y$ for all $w \in r(f') \setminus \{x, z\}$. Then, since $f'$ is a dictatorship of agent $j$, $O_i((\hat{R}_j, R_{-\{i, j\}}), {\cal R}'_i) = x$. By Lemma \ref{range0}, $O_i((\hat{R}_j, R_{-\{i, j\}}), {\cal R}_i) \in \{\{x\}, \{x, y\}\}$ and, thus, we have that $f(R'_i, \hat{R}_j, R_{-\{i, j\}}) \in \{x, y\}$. If $f(R'_i, \hat{R}_j, R_{-\{i, j\}}) = y$, consider any preference $R'_j \in {\cal R}_j$ such that $z \, P'_j \, v$ for all $v \in r(f') \setminus \{z\}$. 
Since agent $j$ is a dictator at $f'$ and since $(R'_j, R_{-j}) \in {\cal D}'$, $O_i((R'_j, R_{-\{i,j\}}), {\cal R}'_i)=z$. 
Then, by Lemma \ref{range0}, $O_i((R'_j, R_{-\{i,j\}}), {\cal R}_i)=z$. 
So, $f(R'_{\{i, j\}}, R_{-\{i,j\}})=z$. 
Agent $j$ then manipulates $f$ at $(R'_i, \hat{R}_j, R_{-\{i, j\}})$ via $R'_j$.
We can therefore conclude that $f(R'_i, \hat{R}_j, R_{-\{i, j\}}) = x$. 
If $x \, P_j \, y$, then agent $j$ manipulates $f$ at $(R'_i, R_{-i})$, where the outcome is $y$, via $\hat{R}_j$. 
Therefore, $y \, P_j \, x$.
Since $(y,x) \not \in S({\cal R}_j)$ by assumption, $\{x, y\} \in S^{-1}({\cal R}_j)$. 
We divide the proof depending on whether or not for all $w \in r(f') \setminus \{x\}$, $(w, y) \not\in S({\cal R}_j)$.

\smallskip

\noindent If $(w, y) \not\in S({\cal R}_j)$ for all $w \in r(f') \setminus \{x\}$, then there exists a preference $R''_j \in {\cal R}_j$ such that $y \, P''_j \, z \, P''_j \, w$ for all $w \in r(f') \setminus \{z\}$.
Since agent $j$ is a dictator at $f'$ and $(R''_j, R_{-j}) \in {\cal D}'$, $O_i((R''_j, R_{-\{i,j\}}), {\cal R}'_i)=z$. 
Then, by Lemma \ref{range0}, $O_i((R''_j, R_{-\{i,j\}}), {\cal R}_i)=z$. 
This implies that $f(R''_j, R'_i, R_{-\{i,j\}})=z$. 
Agent $j$ then manipulates $f$ at this profile via $R_j$ to obtain $y$, which contradicts that $f$ is SP.

\smallskip

\noindent If $(w, y) \in S({\cal R}_j)$ for some $w \in r(f') \setminus \{x\}$, then there is a preference $R''_j \in {\cal R}_j$ such that $w \, P''_j \, v$ for all $v \in r(f) \setminus \{w\}$. 
Since agent $j$ is a dictator at $f'$ and since $(R''_j, R_{-j}) \in {\cal D}'$, $O_i((R''_j, R_{-\{i,j\}}), {\cal R}'_i) = w$. 
Then, by Lemma \ref{range0}, $O_i((R''_j, R_{-\{i,j\}}), {\cal R}_i) = w$. 
This implies that $f(R''_j, R'_i, R_{-\{i,j\}}) = w$. 
Since $(w, y) \in S({\cal R}_j)$ by assumption, $w \, P_j \, y$. 
Agent $j$ thus manipulates $f$ at $(R'_i, R_{-i})$, where the outcome is $y$, via $R''_j$. 
This again contradicts that $f$ is SP.

\end{enumerate}
\end{enumerate}

\noindent This concludes the proof of PART 2.
\bigskip

\noindent {\it PART 3: Suppose that $f'$ is ND and that $|r(f')|=2$.}

\medskip

\noindent By Corollary \ref{range}, $r(f') = \{x, z\}$ and $r(f) = \{x, y, z\}$. The proof is essentially based on the observation that under SP, there is no subprofile $R_{-i} \in {\cal D}_{-i}$ such that $O_i(R_{-i}, {\cal R}'_i) = r(f')$ and $O_i(R_{-i}, {\cal R}_i) = r(f)$.
That is, there is no subprofile $R_{-i} \in {\cal D}_{-i}$ for which agent $i$ ``\emph{acts as a dictator}" for both preference domains ${\cal R}'_i$ and ${\cal R}_i$.
To arrive at this result, which is shown in Lemma \ref{proposition}, we have to establish first a total of five lemmas. 
The first lemma operates under the assumption that there is a subprofile $R_{-i}$ such that the option set of agent $i$ changes when her preference domain is amplified.
We then know from Lemma \ref{range0} that $x$ belongs to the option of agent $i$ in her smaller preference domain and that $y$ enters the option set of agent $i$ in the amplified preference domain. 
In particular, we analyze in this first lemma of the sequence what happens when another agent $j$ deviates from $R_j$ to any other preference. 
It is shown that $x$ still belongs to the option set of agent $i$ in the smaller preference domain; that is, this option set cannot contain only $z$.

\begin{lemma}
\label{xesta}
Consider any $R_{-i} \in {\cal D}_{-i}$ such that $O_i(R_{-i}, {\cal R}'_i) \neq O_i(R_{-i}, {\cal R}_i)$. Then, for all agents $j \neq i$ and all preferences $\hat{R}_j \in {\cal R}_j$, $x \in O_i((\hat{R}_j, R_{-\{i, j\}}), {\cal R}'_i)$.
\end{lemma}

\begin{proof}
Consider any $R_{-i} \in {\cal D}_{-i}$ such that $O_i(R_{-i}, {\cal R}'_i) \neq O_i(R_{-i}, {\cal R}_i)$ and suppose by contradiction that there is an agent $j \neq i$ and a preference $\hat{R}_j \in {\cal R}_j$ such that $x \not\in O_i((\hat{R}_j, R_{-\{i, j\}}), {\cal R}'_i)$. 
Then, since $r(f') = \{x, z\}$, $O_i((\hat{R}_j, R_{-\{i, j\}}), {\cal R}'_i) = z$. 
Since $O_i(R_{-i}, {\cal R}'_i) \neq O_i(R_{-i}, {\cal R}_i)$ by assumption, we have by Lemma \ref{range0} that $x \in O_i(R_{-i}, {\cal R}'_i)$ and $y \in O_i(R_{-i}, {\cal R}_i)$. 
Consider a preference $R_i \in {\cal R}'_i$ such that for all $w \in X \setminus \{x\}$ with $\{x, w\} \in S^{-1}({\cal R}'_i)$, $x \, P_i \, w$. Consider also any $R'_i  \in {\cal R}_i \setminus {\cal R}'_i$ such that for all $v \in X \setminus \{x\}$ with $\{y, v\} \in S^{-1}({\cal R}_i)$, $y \, P'_i \, v$. By Lemma \ref{free}, $x = \max\{R_i|_{O_i(R_{-i}, {\cal R}'_i)}\}$ and $y = \max\{R'_i|_{O_i(R_{-i}, {\cal R}_i)}\}$. Then, by Lemma \ref{maximal}, $f(R) = x$ and $f(R'_i, R_{-i}) = y$. Similarly, since $O_i((\hat{R}_j, R_{-\{i, j\}}), {\cal R}'_i) = z$, we have by Lemma \ref{range0} that $O_i((\hat{R}_j, R_{-\{i, j\}}), {\cal R}_i) = z$. 
Therefore, $f(\hat{R}_j, R_{-j}) = f(R'_i, \hat{R}_j, R_{-\{i, j\}}) = z$. 
Since $f(R'_i, R_{-i}) = y$ and $f(R'_i, \hat{R}_j, R_{-\{i, j\}}) = z$, we have that $y, z \in O_j((R'_i, R_{-\{i, j\}}), {\cal R}_j)$.
And since $f(R) = x$ and $f(\hat{R}_j, R_{-j}) = z$, we have that $x, z \in O_j(R_{-j}, {\cal R}_j)$. 
Then, by Lemma \ref{free}, $\{y, z\}, \{x, z\} \in S^{-1}({\cal R}_j)$. 
We divide the proof depending on whether or not $(x, y) \in S({\cal R}_j)$.

\begin{enumerate}

\item [1.] \emph{Suppose that $(x, y) \not\in S({\cal R}_j)$.}\smallskip

It follows from $\{x,z\}, \{y, z\} \in S^{-1}({\cal R}_j)$ and $(x, y) \not\in S({\cal R}_j)$ that there is a preference $R'_j \in {\cal R}_j$ such that $y \, P'_j \, z \, P'_j \, x$.
Since $x, z \in O_j(R_{-j}, {\cal R}_j)$, $(R'_j, R_{-j}) \in {\cal D}'$, and $r(f') = \{x, z\}$, we can conclude that $O_j(R_{-j}, {\cal R}_j) = \{x, z\}$. 
Then, by Lemma \ref{maximal}, $f(R'_j, R_{-j}) = z$.
Hence, $z \in O_i((R'_j, R_{-\{i,j\}}), {\cal R}'_i)$.  

\noindent We show next that $O_i((R'_j, R_{-\{i,j\}}), {\cal R}'_i) = z$. 
Suppose by contradiction that $O_i((R'_j,$ $R_{-\{i,j\}}), {\cal R}'_i) \neq z$.
Since we have already shown that $z \in O_i((R'_j, R_{-\{i,j\}}), {\cal R}'_i)$, we can conclude that $O_i((R'_j, R_{-\{i,j\}}), {\cal R}'_i) = \{x, z\}$. 
By Lemma \ref{free}, $\{x, z\} \in S^{-1}({\cal R}'_i)$ and, then, by the construction of $R_i \in {\cal R}'_i$, $x \, P_i \, z$. 
Since $O_i((R'_j, R_{-\{i,j\}}), {\cal R}'_i) = \{x, z\}$, there is a preference $\bar{R}_i \in {\cal R}'_i$ such that $f(\bar{R}_i, R'_j, R_{-\{i, j\}}) = x$. 
Agent $i$ thus manipulates $f$ at $(R'_j, R_{-j})$, where the outcome is $z$, via $\bar{R}_i$. 
This contradicts that $f$ is SP. 
Thus, $O_i((R'_j, R_{-\{i,j\}}), {\cal R}'_i) = z$.

\noindent It follows from $O_i((R'_j, R_{-\{i, j\}}), {\cal R}'_i) = z$ and Lemma \ref{range0} that $O_i((R'_j, R_{-\{i, j\}}), {\cal R}_i) = z$. 
Thus, $f(R'_{\{i,j\}}, R_{-\{i, j\}}) = z$. 
Agent $j$ then manipulates $f$ at this profile via $R_j$ to obtain $y$. 
This again contradicts that $f$ is SP.
 
\item [2.] \emph{Suppose that $(x, y) \in S({\cal R}_j)$.}\smallskip

It follows from $\{x, z\}, \{y, z\} \in S^{-1}({\cal R}_j)$ and $(x, y) \in S({\cal R}_j)$ that there is a preference $R'_j \in {\cal R}_j$ such that $x \, P'_j \, z \, P'_j \, y$. 
Note that $(R'_j, R_{-j}) \in {\cal D}'$ and $r(f') = \{x, z\}$, which implies that $f(R'_j, R_{-j}) \in \{x, z\}$. 
If $f(R'_j, R_{-j}) = z$, agent $j$ manipulates $f$ at this profile via $R_j$ to obtain $x$. 
Thus, $f(R'_j, R_{-j}) = x$.

Since $r(f) = \{x,y,z\}$, $f(R'_{\{i, j\}}, R_{-\{i, j\}}) \in \{x,y,z\}$. 
Also observe that $x \, P_j \, y$, because $(x, y) \in S({\cal R}_j)$ by assumption. 
If $f(R'_{\{i, j\}}, R_{-\{i, j\}}) = x$, then agent $j$ manipulates $f$ at $(R'_i, R_{-i})$, where the outcome is $y$, via $R'_j$. 
If $f(R'_{\{i, j\}}, R_{-\{i, j\}}) = y$, then agent $j$ manipulates $f$ at this profile via $\hat{R}_j$ to obtain $z$. 
Hence, $f(R'_{\{i, j\}}, R_{-\{i, j\}}) = z$.

Since $f(R'_j, R_{-j}) = x$ and $f(R'_{\{i, j\}}, R_{-\{i, j\}}) = z$, we have $x,z \in O_i((R'_j,R_{-\{i,j\}},{\cal R}_i)$. 
Then, $\{x, z\} \in S^{-1}({\cal R}_i)$ by Lemma \ref{free}. 
We divide the remainder of the proof depending on whether or not $(z, y) \in S({\cal R}_i)$.

\begin{itemize}
\item [a.] \emph{Suppose that $(z, y) \not \in S({\cal R}_i)$.}\smallskip

Since $\{x,y\},\{x, z\} \in S^{-1}({\cal R}_i)$ and $(z, y) \not \in S({\cal R}_i)$, there is a preference $R''_i \in {\cal R}_i \setminus {\cal R}'_i$ such that $y \, P''_i \, x \, P''_i \, z$.
It follows then from $r(f) = \{x, y, z\}$, $y \in O_i(R_{-i}, {\cal R}_i)$, and Lemma \ref{maximal} that $f(R''_i, R_{-i}) = y$.  
It also follows from $O_i((\hat{R}_j, R_{-\{i, j\}}), {\cal R}_i) = z$ that $f(R''_i, \hat{R}_j, R_{-\{i, j\}}) = z$.

Let us consider $f(R''_i, R'_j, R_{-\{i, j\}})$. 
If $f(R''_i, R'_j, R_{-\{i, j\}}) = x$, then agent $j$ manipulates $f$ at $(R''_i, R_{-i})$, where the outcome is $y$, via $R'_j$ (remember that $(x, y) \in S({\cal R}_j)$ by assumption and, thus, $x \, P_j \, y$).
If $f(R''_i, R'_j, R_{-\{i, j\}}) = y$, then agent $j$ manipulates $f$ at this profile via $\hat{R}_j$ to obtain $z$.
If $f(R''_i, R'_j, R_{-\{i, j\}}) = z$, then agent $i$ manipulates $f$ at this profile via $R_i$ to obtain $x$.
Since $r(f) = \{x, y, z\}$ and $f$ is SP, this is a contradiction.

\item [b.] \emph{Suppose that $(z, y) \in S({\cal R}_i)$.}\smallskip

Since $\{x, z\} \in S^{-1}({\cal R}_i)$ and $(z, y) \in S({\cal R}_i)$, we have that $\{x, z\} \in S^{-1}({\cal R}'_i)$ and $(x, y), (z, y) \in S({\cal R}'_i)$. Therefore, there is a preference $R''_i \in {\cal R}'_i$ such that $z \, P''_i \, x \, P''_i \, y$.
Since $r(f') = \{x,z\}$ and $(R''_i, R'_j, R_{-\{i, j\}}) \in {\cal D}'$, $f(R''_i, R'_j, R_{-\{i, j\}})  \in \{x,z\}$.
If $f(R''_i, R'_j, R_{-\{i, j\}}) = x$, then agent $i$ manipulates $f$ at this profile via $R'_i$ to obtain $z$.
Thus, $f(R''_i, R'_j, R_{-\{i, j\}}) = z$.

\noindent Let us consider $f(R''_i, R_{-i})$. 
If, on the one hand, $f(R''_i, R_{-i}) = z$, then agent $i$ manipulates $f$ at $(R'_i, R_{-i})$, where the outcome is $y$, via $R''_i$ (remember that $(z, y) \in S({\cal R}_i)$ by assumption and, thus, $z \, P'_i \, y$). 
If, on the other hand, $f(R''_i, R_{-i}) = x$, then agent $j$ manipulates $f$ at $(R''_i, R'_j, R_{-\{i, j\}})$, where the outcome is $z$, via $R_j$. 
Since $r(f') = \{x, z\}$ and $(R''_i, R_{-i}) \in {\cal D}'$, this contradicts that $f$ is SP. 
\end{itemize}
\end{enumerate}
This concludes the proof of Lemma \ref{xesta}.
\end{proof}

\noindent We concluded in the former lemma that $x \in O_i((\hat{R}_j, R_{-\{i, j\}}), {\cal R}'_i)$.  
Under the same assumptions, we show next that if $O_i((\hat{R}_j, R_{-\{i, j\}}), {\cal R}'_i)=\{x,z\}$, then $\{y, z\}$ is a free pair for agent $i$.

\begin{lemma}
\label{partialdictator0}
Let $R_{-i} \in {\cal D}_{-i}$ be such that $O_i(R_{-i}, {\cal R}'_i) \neq O_i(R_{-i}, {\cal R}_i)$. If there exists an agent $j \neq i$ and a preference $\hat{R}_j \in {\cal R}_j$ such that $O_i((\hat{R}_j, R_{-\{i, j\}}), {\cal R}'_i) = \{x, z\}$, $\{y, z\} \in S^{-1}({\cal R}'_i)$.
\end{lemma}

\begin{proof}
Consider any $R_{-i} \in {\cal D}_{-i}$ such that $O_i(R_{-i}, {\cal R}'_i) \neq O_i(R_{-i}, {\cal R}_i)$ and take any agent $j \neq i$ and any preference $\hat{R}_j \in {\cal R}_j$ for which $O_i((\hat{R}_j, R_{-\{i, j\}}), {\cal R}'_i) = \{x, z\}$. 
Then, by Lemma \ref{free}, we have that $\{x, z\} \in S^{-1}({\cal R}'_i)$. 
We first show that $(y, z) \not\in S({\cal R}'_i)$. 
Suppose by contradiction that $(y, z) \in S({\cal R}'_i)$.
Since $(x, y) \in S({\cal R}'_i)$ by construction, $(x,z) \in S({\cal R}'_i)$ by transitivity.
This contradicts that $\{x, z\} \in S^{-1}({\cal R}'_i)$. 

\medskip

\noindent We show next that $(z, y) \not\in S({\cal R}'_i)$. 
Suppose by contradiction that $(z,y) \in S({\cal R}'_i)$. 
Since $\{x, z\} \in S^{-1}({\cal R}'_i)$ and $(x,y),(z,y) \in S({\cal R}'_i)$, we have that for each $R''_i \in {\cal R}'_i,$ either $x \, P''_i \, z \, P''_i \, y$ or $z \, P''_i \, x \, P''_i \, y$. 
Let $R^1_i, R^2_i \in {\cal R}'_i$ be such that $x \, P^1_i \, z \, P^1_i \, y$ and $z \, P^2_i \, x \, P^2_i \, y$. 
It follows then from $O_i((\hat{R}_j, R_{-\{i, j\}}), {\cal R}'_i) =\{x,z\}$ and Lemma \ref{maximal} that $f(R^1_i,\hat{R}_j,R_{-\{i,j\}})=x$ and $f(R^2_i,\hat{R}_j,R_{-\{i,j\}})=z$. 

\medskip

\noindent Consider now the preferences in ${\cal R}_i \setminus {\cal R}'_i$. 
Since $O_i(R_{-i}, {\cal R}'_i) \neq O_i(R_{-i}, {\cal R}_i)$ by assumption, it follows from Lemma \ref{range0} that $y \in O_i(R_{-i}, {\cal R}_i) \setminus O_i(R_{-i}, {\cal R}'_i)$ and, therefore, there is a preference $R'_i \in {\cal R}_i \setminus {\cal R}'_i$ such that $f(R'_i,R_{-i})=y$.
Since $(z,y) \in S({\cal R}'_i)$ by assumption, $(z,y) \in S({\cal R}_i)$ and, then, each $R''_i \in {\cal R}_i \setminus {\cal R}'_i$ is such that $z \, P''_i \, y \, P''_i \, x$. 
Thus, $z \, P'_i \, y \, P'_i \, x$.
If $f(R'_i,\hat{R}_j,R_{-\{i,j\}}) \in \{x,y\}$, then agent $i$ manipulates $f$ at this profile via $R_i^2$ to obtain $z$. 
So, $f(R'_i, \hat{R}_j, R_{-\{i,j\}}) = z$.
Hence, we can conclude that $y,z \in O_j((R'_i, R_{-\{i, j\}}),{\cal R}_j)$. By Lemma \ref{free}, $\{y,z\} \in S^{-1}({\cal R}_j)$.

\medskip

\noindent Consider now $f(R_i^2, R_{-i})$. If $f(R_i^2, R_{-i})=z$, then agent $i$ manipulates $f$ at $(R'_i, R_{-i})$, where the outcome is $y$, via $R_i^2$. Then, since $(R_i^2, R_{-i}) \in {\cal D}'$ and $r(f') = \{x, z\}$, $f(R_i^2, R_{-i})=x$. 
Since $f(R_i^2, R_{-i})=x$ and $f(R_i^2, \hat{R}_j, R_{-\{i,j\}})=z$, we have that $x,z \in O_j((R_i^2, R_{-\{i, j\}}),{\cal R}_j)$. 
By Lemma \ref{free}, $\{x,z\} \in S^{-1}({\cal R}_j)$. 
Consider finally $f(R_i^1, R_{-i})$. If $f(R_i^1, R_{-i})=z$, then agent $i$ manipulates $f$ at this profile via $R_i^2$ to obtain $x$. 
Since $(R_i^1, R_{-i}) \in {\cal D}'$ and $r(f') = \{x, z\}$, we have that $f(R_i^1, R_{-i})=x$. 
We further divide the case depending on whether or not $(x, y)$ belongs to $S({\cal R}_j)$.

\begin{enumerate}
\item[a.] \emph{Suppose that $(x,y) \in S({\cal R}_j)$}.

Since $(x,y) \in S({\cal R}_j)$ by assumption and $\{x, z\}, \{y, z\} \in S^{-1}({\cal R}_j)$, there is a preference $R'_j \in {\cal R}_j$ such that $x \, P'_j \, z \, P'_j \, y$. Given that $(R_i^2, R'_j, R_{-\{i, j\}}) \in {\cal D}'$ and $r(f') = \{x, z\}$, $f(R_i^2, R'_j, R_{-\{i, j\}}) \in \{x, z\}$. 
If $f(R_i^2,R'_j,R_{-\{i, j\}}) =z$, then agent $j$ manipulates $f$ at this profile via $R_j$ to obtain $x$. 
Thus, $f(R_i^2, R'_j, R_{-\{i, j\}}) = x$. 
Consider now $f(R'_{\{i,j\}}, R_{-\{i, j\}})$. 
Since $r(f) = \{x, y, z\}$, $f(R'_{\{i,j\}}, R_{-\{i, j\}}) \in \{x, y, z\}$. 
If $f(R'_{\{i,j\}},R_{-\{i, j\}}) = x$, then agent $j$ manipulates $f$ at $(R'_i, R_{-i})$, where the outcome is $y$, via $R'_j$ (remember that $(x, y) \in S({\cal R}_j)$ by assumption and, thus, $x \, P_j \, y$). 
If $f(R'_{\{i,j\}},R_{-\{i, j\}}) = y$, then agent $j$ manipulates $f$ at this profile via $\hat{R}_j$ to obtain $z$. 
If $f(R'_{\{i,j\}}, R_{-\{i, j\}}) = z$, then agent $i$ manipulates $f$ at $(R_i^2, R'_j, R_{-\{i, j\}})$, where the outcome is $x$, via $R'_i$. 
This contradicts that $f$ is SP.

\item [b.] \emph{Suppose that $(x, y) \not\in S({\cal R}_j)$}.

Since $(x, y) \not\in S({\cal R}_j)$ by assumption and $\{x, z\}, \{y, z\} \in S^{-1}({\cal R}_j)$, there is a preference $R'_j \in {\cal R}_j$ such that $y \, P'_j \, z \, P'_j \, x$. Given that $(R_i^2, R'_j, R_{-\{i,j\}}) \in {\cal D}'$ and $r(f') = \{x, z\}$, $f(R_i^2, R'_j, R_{-\{i, j\}}) \in \{x, z\}$.
If $f(R_i^2, R'_j, R_{-\{i,j\}}) = x$, then agent $j$ manipulates $f$ at this profile via $\hat{R}_j$ to obtain $z$. 
Thus, $f(R_i^2, R'_j, R_{-\{i,j\}}) = z$.
Consider now $f(R'_{\{i,j\}}, R_{-\{i,j\}})$. 
If $f(R'_{\{i,j\}}, R_{-\{i,j\}}) \in \{x,y\}$, then agent $i$ manipulates $f$ at this profile via $R_i^2$ to obtain $z$. 
If $f(R'_{\{i,j\}}, R_{-\{i,j\}}) = z$, then agent $j$ manipulates $f$ at this profile via $R_j$ to obtain $y$. 
Since $r(f) = \{x, y, z\}$, this contradicts that $f$ is SP.
\end{enumerate}
This concludes the proof of Lemma \ref{partialdictator0}.
\end{proof}

\noindent Observe that under the assumptions of Lemma \ref{partialdictator0} (the existence of a subprofile $R_{-i} \in {\cal D}_{-i}$ for which the option set for agent $i$ changes when her domain is amplified from ${\cal R}'_i$ to ${\cal R}_i$ and the existence of an alternative preference for some agent $j$ such that, modifying only this preference, the option set for agent $i$ is $\{x, z\}$ under ${\cal R}'_i$), we can prove that $\{x, z\}, \{y, z\} \in S^{-1}({\cal R}_i)$.
In fact, it follows from the fact that $\{x, z\}$ is the option set of agent $i$ in some subprofile and Lemma \ref{free} that $\{x, z\} \in S^{-1}({\cal R}_i)$.
And we know from Lemma \ref{partialdictator0} that $\{y, z\} \in S^{-1}({\cal R}_i)$. 
Consider then any preferences $R_i^1, R_i^2, \ldots, R_i^6$ such that $x \, P_i^1 \, y \, P_i^1 \, z$, $x \, P_i^2 \, z \, P_i^2 \, y$, $z \, P_i^3 \, x \, P_i^3 \, y$, $z \, P_i^4 \, y \, P_i^4 \, x$, $y \, P_i^5 \, z \, P_i^5 \, x$, $y \, P_i^6 \, x \, P_i^6 \, z$, and $v \, P_l \, w$ for all $(v, w) \in S({\cal R}_i)$ and all $l \in \{1, 2, \ldots, 6\}$.  
Then, under the assumptions explained above, $R_i^1, R_i^2, R_i^3 \in {\cal R}'_i$ and $R_i^4,R_i^5,R_i^6 \in {\cal R}_i \setminus {\cal R}'_i$. 
We will make use of these preferences throughout the proofs of the next two lemmas.

\medskip

\noindent Under the same assumptions as before, we show that there is a preference $R'_j \in {\cal R}_j$ for agent $j$ so that the option set of agent $i$ is, for both her original and the amplified preference domain, equal to the whole range. So, the lemma establishes that agent $i$ ``\emph{acts as a dictator}" at subprofile $(R'_j, R_{\{i,j\}})$ for both ${\cal R}'_i$ and ${\cal R}_i$. 

\begin{lemma}
\label{existencia2}
Let $R_{-i} \in {\cal D}_{-i}$ be such that $O_i(R_{-i}, {\cal R}'_i) \neq O_i(R_{-i}, {\cal R}_i)$. If there exists an agent $j \neq i$ and a preference $\hat{R}_j \in {\cal R}_j$ such that $O_i((\hat{R}_j, R_{-\{i, j\}}), {\cal R}'_i) = \{x, z\}$, then there is a preference $R'_j \in {\cal R}_j$ such that $O_i((R'_j, R_{-\{i, j\}}), {\cal R}'_i) = r(f')$ and $O_i((R'_j, R_{-\{i, j\}}), {\cal R}_i) = r(f)$.
\end{lemma}

\begin{proof}
Consider any $R_{-i} \in {\cal D}_{-i}$ such that $O_i(R_{-i}, {\cal R}'_i) \neq O_i(R_{-i}, {\cal R}_i)$ and take any agent $j \neq i$ and any preference $\hat{R}_j \in {\cal R}_j$ such that $O_i((\hat{R}_j, R_{-\{i, j\}}), {\cal R}'_i) = \{x, z\}$. Note that, since $O_i(R_{-i}, {\cal R}'_i) \neq O_i(R_{-i}, {\cal R}_i)$, $y \in O_i(R_{-i}, {\cal R}_i)$ by Lemma \ref{range0}.
Suppose, by contradiction, that there is no $R'_j \in {\cal R}_j$ such that  $O_i((R'_j, R_{-\{i, j\}}), {\cal R}'_i) = r(f') = \{x,z\}$ and $O_i((R'_j, R_{-\{i, j\}}), {\cal R}_i) = r(f) = \{x,y,z\}$. 
If $O_i(R_{-i}, {\cal R}'_i) = \{x, z\}$, then $O_i(R_{-i}, {\cal R}'_i) = r(f')$ and $O_i(R_{-i}, {\cal R}_i) = r(f)$ by Lemma \ref{range0} and we reach a contradiction by setting $R'_j$ equal to $R_j$. 
Consequently, $O_i(R_{-i}, {\cal R}'_i) \neq \{x, z\}$. 
Thus, by Lemma \ref{range0}, $O_i(R_{-i}, {\cal R}'_i) = x$ and $O_i(R_{-i}, {\cal R}_i) = \{x, y\}$. 
Lemma \ref{maximal} then implies that $f(R_i^3,R_{-i})=x$ and $f(R_i^6,R_{-i})=y$. 
Also, if $O_i((\hat{R}_j, R_{-\{i, j\}}), {\cal R}_i) = r(f)$, then we reach a contradiction by setting $R'_j$ equal to $\hat{R}_j$. 
Consequently, $O_i((\hat{R}_j, R_{-\{i, j\}}), {\cal R}_i) = r(f') = \{x, z\}$. 
Lemma \ref{maximal} then implies that $f(R^3_i,\hat{R_j},R_{-\{i,j\}})=z$ and $f(R_i^6, \hat{R}_j, R_{-\{i,j\}})=x$.

\medskip

\noindent It follows from $f(R^3_i,R_{-i})=x$ and $f(R^3_i,\hat{R_j},R_{-\{i,j\}})=z$ that $x, z \in O_j((R_i^3, R_{-\{i, j\}}), {\cal R}_j)$. 
Hence, by Lemma \ref{free}, $\{x,z\} \in S^{-1}({\cal R}_j)$. 
Moreover, it follows from $f(R_i^6, R_{-i})=y$ and $f(R_i^6,\hat{R}_j, R_{-\{i,j\}})=x$ that $x, y \in O_j((R_i^6, R_{-\{i, j\}}), {\cal R}_j)$. Hence, by Lemma \ref{free}, $\{x,y\} \in S^{-1}({\cal R}_j)$. 
Since $\{x,z\},\{x,y\}  \in S^{-1}({\cal R}_j)$, consider any $\bar{R}_j \in {\cal R}_j$ such that $y \, \bar{P}_j \, x$ and $z \, \bar{P}_j \, x$.

\medskip

\noindent We start by showing that $O_i((\bar{R}_j, R_{-\{i, j\}}), {\cal R}'_i) = r(f') = \{x,z\}$. By Lemma \ref{xesta}, $x \in O_i((\bar{R}_j, R_{-\{i, j\}}), {\cal R}'_i)$. Hence, it only remains to be shown that $z \in O_i((\bar{R}_j, R_{-\{i, j\}}), {\cal R}'_i)$. Suppose otherwise, that is, $z \not\in O_i((\bar{R}_j, R_{-\{i, j\}}), {\cal R}'_i)$. It must then be the case that $O_i((\bar{R}_j, R_{-\{i, j\}}), {\cal R}'_i) = x$. This implies that $f(R_i^3, \bar{R}_j, R_{-\{i, j\}}) = x$. It also follows from $O_i((\hat{R}_j, R_{-\{i, j\}}), {\cal R}'_i) = \{x, z\}$ and Lemma \ref{maximal} that $f(R_i^3, \hat{R}_j, R_{-\{i, j\}}) = z$. Agent $j$ thus manipulates $f$ at $(R_i^3, \bar{R}_j, R_{-\{i, j\}})$ via $\hat{R}_j$. Since this contradicts that $f$ is SP, we conclude that $O_i((\bar{R}_j, R_{-\{i, j\}}), {\cal R}'_i) = r(f')$.

\medskip

\noindent We show next that $O_i((\bar{R}_j, R_{-\{i, j\}}), {\cal R}_i) = r(f)$. By assumption, $r(f)=\{x,y,z\}$. Since ${\cal R}'_i \subset {\cal R}_i$ and $O_i((\bar{R}_j, R_{-\{i, j\}}), {\cal R}'_i) = \{x, z\}$, it only remains to be shown that $y \in O_i((\bar{R}_j, R_{-\{i, j\}}), {\cal R}_i)$. Suppose otherwise, that is, $y \not \in O_i((\bar{R}_j, R_{-\{i, j\}}), {\cal R}_i)$. Consequently, $O_i((\bar{R}_j, R_{-\{i, j\}}), {\cal R}_i)=\{x, z\}$. Then, by Lemma \ref{maximal}, $f(R_i^6, \bar{R}_j, R_{-\{i, j\}}) = x$. It also follows from $y \in O_i(R_{-i}, {\cal R}_i)$, $r(f) = \{x, y, z\}$, and Lemma \ref{maximal} that $f(R_i^6, R_{-i}) = y$. Agent $j$ thus manipulates $f$ at $(R_i^6, \bar{R}_j, R_{-\{i, j\}})$ via $R_j$. Since this again contradicts that $f$ is SP, we conclude that $O_i((\bar{R}_j, R_{-\{i, j\}}), {\cal R}_i) = r(f)$. Then, we reach a contradiction by setting $R'_j$ equal to $\bar{R}_j$. 
\end{proof}

\noindent To introduce the next lemma, we need the following definition. 
We say that two preferences $R_j,R'_j \in {\cal R}_j$ are adjacent whenever there is an alternative $v \in \{x,y,z\}$ and another alternative $w \in \{x,y,z\} \setminus \{v\}$ such that $v \, P_j \, w$, $w \, P'_j \, v$, and for all pairs of alternatives $\{s,t\} \subset \{x, y, z\}$ such that $\{s, t\} \neq \{v, w\}$, $s \, P_j \, t \Leftrightarrow s \, P'_j \, t$. 
Informally, two preferences are adjacent if they only differ in one binary comparison on the set of alternatives $\{x,y,z\}$. 

\medskip

\noindent Maintaining the same assumptions as before (\emph{i.e.}, that the option set for agent $i$ changes when her preference domain is amplified at the subprofile $R_{-i} \in {\cal D}_{-i}$, and that a deviation of agent $j$ can provoke that the option set for agent $i$ in her smaller domain is the whole range of $f'$), we establish next that if there is some alternative preference $R'_j \in {\cal R}_j$ so that agent $i$ ``\emph{acts as a dictator}" at the subprofile $(R'_j,R_{-\{i,j\}})$ for her smaller and her amplified preference domains, then agent $i$ remains to ``\emph{acts as a dictator}" for both preference domains whenever agent $j$ has a preference $R''_j$ that is adjacent of $R'_j$.

\begin{lemma}
\label{extension}
Let $R_{-i} \in {\cal D}_{-i}$ be such that $O_i(R_{-i}, {\cal R}'_i) \neq O_i(R_{-i}, {\cal R}_i)$. If there exists an agent $j \neq i$ and preferences $\hat{R}_j, R'_j, R''_j \in {\cal R}_j$ such that $O_i((\hat{R}_j, R_{-\{i, j\}}), {\cal R}'_i) = \{x, z\}$, such that $R'_j$ and $R''_j$ are adjacent, and such that $O_i((R'_j, R_{-\{i, j\}}), {\cal R}'_i) = r(f')$ and $O_i((R'_j, R_{-\{i, j\}}), {\cal R}_i) = r(f)$, then $O_i((R''_j, R_{-\{i, j\}}), {\cal R}'_i) = r(f')$ and $O_i((R''_j, R_{-\{i, j\}}), {\cal R}_i) = r(f)$.
\end{lemma}

\begin{proof}
Consider any $R_{-i} \in {\cal D}_{-i}$ such that $O_i(R_{-i}, {\cal R}'_i) \neq O_i(R_{-i}, {\cal R}_i)$ and take any agent $j \neq i$ and any three preferences $\hat{R}_j, R'_j, R''_j \in {\cal R}_j$ such that $R'_j$ and $R''_j$ are adjacent, such that $O_i((\hat{R}_j, R_{-\{i, j\}}), {\cal R}'_i) = O_i((R'_j, R_{-\{i, j\}}), {\cal R}'_i) = r(f')= \{x, z\}$, and such that $O_i((R'_j, R_{-\{i, j\}}), {\cal R}_i)=r(f)=\{x,y,z\}$. 
Then, by Lemma \ref{maximal}, $f(R^3_i, R'_j, R_{-\{i, j\}})=f(R^4_i, R'_j, R_{-\{i, j\}})=z$ and $f(R^5_i, R'_j, R_{-\{i, j\}})=f(R^6_i, R'_j, R_{-\{i, j\}})=y$.

\medskip

\noindent By Lemma \ref{xesta}, $x \in O_i((R''_j, R_{-\{i, j\}}), {\cal R}'_i)$. 
Therefore, $x \in O_i((R''_j, R_{-\{i, j\}}), {\cal R}_i)$ by Lemma \ref{range0}.
Hence, it only remains to be shown that $z \in O_i((R''_j, R_{-\{i, j\}}), {\cal R}'_i)$ and that $y \in O_i((R''_j, R_{-\{i, j\}}), {\cal R}_i)$. 
Since $R'_j$ and $R''_j$ are adjacent by assumption, they must have the same maximal or the same minimal alternative on the set $\{x, y, z\}$ (but not both). Suppose that the maximal elements coincide (the proof of the other case is similar), that is, $\max\{R'_j|_{\{x, y, z\}}\} = \max\{R''_j|_{\{x, y, z\}}\}$. 
There are two cases.

\begin{enumerate}
\item Suppose that $\max\{R'_j|_{\{x, y, z\}}\} = \max\{R''_j|_{\{x, y, z\}}\} \in \{x,z\}$. 
Suppose first by contradiction that $z \not \in O_i((R''_j, R_{-\{i, j\}}), {\cal R}'_i)$. 
Then, $O_i((R''_j, R_{-\{i, j\}}), {\cal R}'_i) = x$. 
Hence, $f(R_i^3, R''_j, R_{-\{i, j\}}) = x$. 
First, if $\max\{R'_j|_{\{x, y, z\}}\} = \max\{R''_j|_{\{x, y, z\}}\} = z$, then
agent $j$ manipulates $f$ at $(R_i^3, R''_j, R_{-\{i, j\}})$ via $R'_j$ to obtain $z$.
Second, if  $\max\{R'_j|_{\{x, y, z\}}\} = \max\{R''_j|_{\{x, y, z\}}\} = x$, then agent $j$ manipulates $f$ at $(R_i^3, R'_j, R_{-\{i, j\}})$, where the outcome is $z$, via $R''_j$.
In any case, this contradicts that $f$ is SP and, thus, $z \in O_i((R''_j, R_{-\{i, j\}}), {\cal R}'_i)$.

Next, suppose by contradiction that $y \not \in O_i((R''_j, R_{-\{i, j\}}), {\cal R}_i)$. Consequently, we have that $O_i((R''_j, R_{-\{i, j\}}), {\cal R}_i) = \{x,z\}$ and, thus, by Lemma \ref{maximal}, $f(R_i^5, R''_j, R_{-\{i, j\}}) = z$ and $f(R_i^6, R''_j, R_{-\{i, j\}}) = x$. First, if $\max\{R'_j|_{\{x, y, z\}}\} = \max\{R''_j|_{\{x, y, z\}}\} = z$, then agent $j$ manipulates $f$ at $(R_i^5, R'_j, R_{-\{i, j\}})$, where the outcome is $y$, via $R''_j$. 
Second, if $\max\{R'_j|_{\{x, y, z\}}\} = \max\{R''_j|_{\{x, y, z\}}\} =  x$, then agent $j$ manipulates $f$ at $(R_i^6, R'_j, R_{-\{i, j\}})$, where the outcome is $y$, via $R''_j$. 
In any case, this contradicts that $f$ is SP and, thus, $y \in O_i((R''_j, R_{-\{i, j\}}), {\cal R}_i)$.

\item Suppose that $\max\{R'_j|_{\{x, y, z\}}\} = \max\{R''_j|_{\{x, y, z\}}\} = y$.
Suppose first by contradiction that $y \not \in O_i((R''_j, R_{-\{i, j\}}), {\cal R}_i)$. Consequently, we have $f(R^5_i, R''_j, R_{-\{i, j\}}) \in \{x,z\}$. Agent $j$ thus manipulates $f$ at this profile via $R'_j$ to obtain $y$. 
This contradicts that $f$ is SP and, thus, $y \in O_i((R''_j, R_{-\{i, j\}}), {\cal R}_i)$.

Next, suppose by contradiction that $z \not \in O_i((R''_j, R_{-\{i, j\}}), {\cal R}'_i)$. Consequently, we have that $O_i((R''_j, R_{-\{i, j\}}), {\cal R}'_i) = x$. Since $y \in O_i((R''_j, R_{-\{i, j\}}), {\cal R}_i)$, it follows from Lemma \ref{range0} that $O_i((R''_j, R_{-\{i, j\}}), {\cal R}_i) = \{x,y\}$. Then, by Lemma \ref{maximal}, $f(R_i^4, R''_j, R_{-\{i, j\}})=y$. Agent $j$ thus manipulates $f$ at $(R_i^4, R'_j, R_{-\{i, j\}})$, where the outcome is $z$, via $R''_j$. 
This again contradicts that $f$ is SP and, thus, $z \in O_i((R''_j, R_{-\{i, j\}}), {\cal R}'_i)$.
\end{enumerate}

\noindent This concludes the proof of the lemma.
\end{proof}

\noindent Under the same assumptions as in all preceding lemmas, it is shown next that agent $i$  ``\emph{acts as a dictator}"  for both of her preference domains at all subprofiles $(R'_j,R_{-\{i,j\}})$ for any $R'_j \in {\cal R}_j$.

\begin{lemma}
\label{pre-proposition}
Let $R_{-i} \in {\cal D}_{-i}$ be such that $O_i(R_{-i}, {\cal R}'_i) \neq O_i(R_{-i}, {\cal R}_i)$. 
If there is an agent $j \neq i$ and a preference $\hat{R}_j \in {\cal R}_j$ such that $O_i((\hat{R}_j, R_{-\{i, j\}}), {\cal R}'_i) = \{x, z\}$, then for all $R'_j \in {\cal R}_j$, $O_i((R'_j, R_{-\{i, j\}}), {\cal R}'_i) = r(f')$ and $O_i((R'_j, R_{-\{i, j\}}), {\cal R}_i) = r(f)$.
\end{lemma}

\begin{proof}
Consider any $R_{-i} \in {\cal D}_{-i}$ such that $O_i(R_{-i}, {\cal R}'_i) \neq O_i(R_{-i}, {\cal R}_i)$ and take any agent $j \neq i$ and any preference $\hat{R}_j \in {\cal R}_j$ such that $O_i((\hat{R}_j, R_{-\{i, j\}}), {\cal R}'_i) = r(f')= \{x, z\}$. By Lemma \ref{existencia2}, there is a preference $R''_j \in {\cal R}_j$ such that $O_i((R''_j, R_{-\{i, j\}}), {\cal R}'_i) = r(f')$ and $O_i((R''_j, R_{-\{i, j\}}), {\cal R}_i) = r(f)$. 
Observe next that any non-conditional domain ${\cal R}_j$ satisfies the following property: for each $\tilde{R}''_j, \tilde{R}'_j \in {\cal R}_j$, there exists a path of preference rankings from $\tilde{R}''_j$ to $\tilde{R}'_j$, $\tilde{R}''_j = \tilde{R}^1_j, \tilde{R}_j^2, \ldots, \tilde{R}_j^n = \tilde{R}'_j$, such that for each $l \in \{1, 2, \ldots, n-1\}$, $\tilde R_j^l \in {\cal R}_j$ and the preferences $\tilde{R}_j^{l}$ and $\tilde{R}_j^{l+1}$ are adjacent.\footnote{The construction of this path is simple: $\tilde{R}_j^l$ is constructed from $\tilde{R}_j^{l-1}$ by changing a binary comparison in which $\tilde{R}_j$ and $\tilde{R}'_j$ differs from the ranking of the pair established in $\tilde{R}_j$ to the ranking of the pair established in $\tilde{R}'_j$.}

\medskip

\noindent Consider any $R'_j \in {\cal R}_j$. 
By the former property, there exists a path of rankings from $R''_j$ to $R'_j$, $R''_j = \tilde{R}^1_j, \tilde{R}_j^2, \ldots, \tilde{R}_j^n = R'_j$, such that for each $l \in \{1, 2, \ldots, n-1\}$, $\tilde{R}_j^l \in {\cal R}_j$, and $\tilde{R}_j^{l}$ and $\tilde{R}_j^{l+1}$ are adjacent. 
The iterative application of Lemma \ref{extension} implies that $O_i((R'_j, R_{-\{i, j\}}), {\cal R}'_i) = r(f')$ and $O_i((R'_j, R_{-\{i, j\}}), {\cal R}_i) = r(f)$.
\end{proof}

\noindent Finally, we are ready to show that there is no subprofile $R_{-i} \in {\cal D}_{-i}$ such that $O_i(R_{-i}, {\cal R}'_i) = \{x, z\}$ and $O_i(R_{-i}, {\cal R}_i) = \{x, y, z\}$.

\begin{lemma}
\label{proposition}
There is no subprofile $R_{-i} \in {\cal D}_{-i}$ such that $O_i(R_{-i}, {\cal R}'_i) = r(f') = \{x, z\}$ and $O_i(R_{-i}, {\cal R}_i) = r(f) = \{x, y, z\}$.
\end{lemma}

\begin{proof}
Suppose by contradiction that there exists a subprofile $R_{-i} \in {\cal D}_{-i}$ such that $O_i(R_{-i}, {\cal R}'_i) = r(f') = \{x, z\}$ and $O_i(R_{-i}, {\cal R}_i) = r(f) = \{x, y, z\}$. 
Consider any agent $j \in N \setminus \{i\}$. 
Since $O_i((R_j,R_{-\{i,j\}}),{\cal R}'_i) = r(f')$ by assumption,  we can apply Lemma \ref{pre-proposition} (with $R_j$ playing the role of $\hat{R}_j$ in the antecedent of the lemma) to see that for all $R'_j \in {\cal R}_j$, $O_i((R'_j, R_{-\{i, j\}}), {\cal R}'_i) = r(f')$ and $O_i((R'_j, R_{-\{i, j\}}), {\cal R}_i) = r(f)$. 

\medskip

\noindent Next, consider any agent $k \in N \setminus  \{i, j\}$. Since we already know that for all $R'_j \in {\cal R}_j$, $O_i((R'_j, R_{-\{i, j\}}), {\cal R}'_i) = r(f')$ and $O_i((R'_j, R_{-\{i, j\}}), {\cal R}_i) = r(f)$, we can apply Lemma \ref{pre-proposition} again (this time with $(R'_j, R_{-\{i, j\}})$ playing the role of $R_{-i}$, with $k$ playing the role of $j$, and with $R_k$ playing the role of $\hat{R}_j$) to see that for all $R'_k \in {\cal R}_k$, $O_i((R'_{\{j,k\}}, R_{-\{i, j,k\}}), {\cal R}'_i) = r(f')$ and $O_i((R'_{\{j,k\}}, R_{-\{i, j,k\}}), {\cal R}_i) = r(f)$. 
Proceeding iteratively in this way we obtain that for all $R'_{-i} \in {\cal D}_{-i}, O_i(R'_{-i}, {\cal R}'_i) = r(f') = \{x, z\}$ and $O_i(R'_{-i}, {\cal R}_i) = r(f) = \{x, y, z\}$. Hence, $f'$ is dictatorial and agent $i$ is the dictator. This contradicts the initial assumption of PART 3 that $f'$ is ND.
\end{proof}

\noindent We now introduce the following notation: for any $i \in N$, any $S \subseteq X$ and any $R, R' \in {\cal D}$, we define $D_S(R_{-i}, R'_{-i}) = \{j \in N \setminus \{i\}: R_j|_S \neq R'_j|_S \}$ and $d_S(R_{-i}, R'_{-i}) = |D_S(R_{-i}, R'_{-i})|$. That is, $D_S(R_{-i}, R'_{-i})$ corresponds to the set of agents who have different preferences on $S$ at $R_{-i}$ and $R'_{-i}$, and $d_S(R_{-i}, R'_{-i})$ is the cardinality of this set.

\medskip

\noindent We are now ready to complete the proof of PART 3. 
By Lemma \ref{proposition}, there is no subprofile $R_{-i} \in {\cal D}_{-i}$ such that $O_i(R_{-i}, {\cal R}'_i) = r(f') = \{x, z\}$ and $O_i(R_{-i}, {\cal R}_i) = r(f) = \{x, y, z\}$. 
It follows then from $r(f')=\{x,z\}$, $r(f)=\{x,y,z\}$, and Lemma \ref{range0} that there are two subprofiles $R'_{-i}, R''_{-i} \in {\cal D}_{-i}$ such that $O_i(R'_{-i}, {\cal R}'_i) = x$, $O_i(R'_{-i}, {\cal R}_i) = \{x, y\}$ and $O_i(R''_{-i}, {\cal R}'_i) = O_i(R''_{-i}, {\cal R}_i) \in \{\{z\}, \{x, z\}\}$. Consider then any two subprofiles $\bar{R}_{-i}, \hat{R}_{-i} \in {\cal D}_{-i}$ such that

\begin{itemize}
\item[(a)] $O_i(\bar{R}_{-i}, {\cal R}'_i) = x$ and $O_i(\bar{R}_{-i}, {\cal R}_i) = \{x, y\}$.

\item[(b)] $O_i(\hat{R}_{-i}, {\cal R}'_i) = O_i(\hat{R}_{-i}, {\cal R}_i) \in \{\{z\}, \{x, z\}\}$.

\item[(c)] For all $\bar R'_{-i}, \hat R'_{-i} \in {\cal D}_{-i}$ such that $O_i(\bar R'_{-i}, {\cal R}'_i) = x$, $O_i(\bar{R}'_{-i}, {\cal R}_i) = \{x, y\}$, and $O_i(\hat R'_{-i}, {\cal R}'_i) = O_i(\hat R'_{-i}, {\cal R}_i) \in \{\{z\}, \{x, z\}\}$, we have  $d_{r(f)}(\bar R'_{-i}, \hat R'_{-i}) \geq d_{r(f)}(\bar{R}_{-i}, \hat{R}_{-i})$.

\item[(d)] For all $j \not\in D_{r(f)}(\bar{R}_{-i}, \hat{R}_{-i})$, $\bar{R}_j = \hat{R}_j$.
\end{itemize}

\noindent It is obvious that two subprofiles with characteristics $(a)$, $(b)$, and $(c)$ exist.
Given that, by SP, the preferences over alternatives outside the range of $f$ cannot alter the outcome of $f$ and, thus, cannot change the option sets, we can assume $(d)$. The next lemma proves that all agents who have different preferences on $r(f)$ at $\bar R_{-i} \in {\cal D}_{-i}$ and $\hat R_{-i} \in {\cal D}_{-i}$ have $\{x,y\}$ and $\{x,z\}$ among their free pairs.

\begin{lemma}
\label{libres}
For all $j \in D_{r(f)}(\bar{R}_{-i}, \hat{R}_{-i})$, $\{x, y\}, \{x, z\} \in S^{-1}({\cal R}_j)$. 
\end{lemma}

\begin{proof}
We first show that for all non-empty subsets of agents $S \subset D_{r(f)}(\bar{R}_{-i}, \hat{R}_{-i})$, $O_i((\hat{R}_S, \bar{R}_{-(S \cup \{i\})}), {\cal R}'_i) = O_i((\hat{R}_S, \bar{R}_{-(S \cup \{i\})}), {\cal R}_i) = x$. 
Suppose by contradiction that there is a non-empty subset $S \subset D_{r(f)}(\bar{R}_{-i}, \hat{R}_{-i})$ such that either $O_i((\hat{R}_S, \bar{R}_{-(S \cup \{i\})}), {\cal R}'_i) \neq x$ or $O_i((\hat{R}_S, \bar{R}_{-(S \cup \{i\})}), {\cal R}_i) \neq x$. 

\begin{itemize}
\item Suppose that $O_i((\hat{R}_S, \bar{R}_{-(S \cup \{i\})}), {\cal R}'_i) \neq x$. 
Since $r(f') = \{x,z\}$, we have that $O_i((\hat{R}_S, \bar{R}_{-(S \cup \{i\})}), {\cal R}'_i) \in \{\{z\},\{x,z\}\}$.
If we have that $O_i((\hat{R}_S, \bar{R}_{-(S \cup \{i\})}), {\cal R}_i) =  O_i((\hat{R}_S, \bar{R}_{-(S \cup \{i\})}), {\cal R}'_i)$, then $d_{r(f)}(\bar{R}_{-i}, (\hat{R}_S, \bar{R}_{-(S \cup \{i\})})) < d_{r(f)}(\bar{R}_{-i}, \hat{R}_{-i})$.
This contradicts (c).
If $O_i((\hat{R}_S, \bar{R}_{-(S \cup \{i\})}), {\cal R}_i) \neq O_i((\hat{R}_S, \bar{R}_{-(S \cup \{i\})}), {\cal R}'_i)$, then, by Lemma \ref{range0}, $O_i((\hat{R}_S, \bar{R}_{-(S \cup \{i\})}), {\cal R}_i) = \{x, z\}$ and $O_i((\hat{R}_S, \bar{R}_{-(S \cup \{i\})}), {\cal R}'_i) = \{x, y, z\}$.
This contradicts Lemma \ref{proposition}.

\item Suppose that $O_i((\hat{R}_S, \bar{R}_{-(S \cup \{i\})}), {\cal R}'_i) = x$ and $O_i((\hat{R}_S, \bar{R}_{-(S \cup \{i\})}), {\cal R}_i) \neq x$. 
By Lemma \ref{range0}, $O_i((\hat{R}_S, \bar{R}_{-(S \cup \{i\})}), {\cal R}_i) =\{x,y\}$.
We have then that $d_{r(f)}((\hat{R}_S, \bar{R}_{-(S \cup \{i\})}), \hat{R}_{-i}) < d_{r(f)}(\bar{R}_{-i}, \hat{R}_{-i})$.
This contradicts (c).

\end{itemize}

\noindent We thus conclude that for all non-empty $S \subset D_{r(f)}(\bar{R}_{-i}, \hat{R}_{-i})$, $O_i((\hat{R}_S, \bar{R}_{-(S \cup \{i\})}), {\cal R}'_i) = O_i((\hat{R}_S, \bar{R}_{-(S \cup \{i\})}), {\cal R}_i) = x$. 
Take any agent $j \in D_{r(f)}(\bar{R}_{-i}, \hat{R}_{-i})$.  
First, setting $S = D_{r(f)}(\bar{R}_{-i}, \hat{R}_{-i}) \setminus \{j\}$, and taking into account that $\bar{R}_k = \hat{R}_k$ for all $k \not\in D_{r(f)}(\bar{R}_{-i}, \hat{R}_{-i})$ by condition (d), we can see that $O_i((\hat{R}_{N \setminus \{i, j\}}, \bar{R}_j), {\cal R}_i) = x$. 
Thus, for all $R_i \in {\cal R}_i$, $f(\hat{R}_{N \setminus \{i, j\}}, \bar{R}_j, R_i) = x$. 
Since $O_i(\hat{R}_{-i}, {\cal R}_i) \in \{\{z\}, \{x, z\}\}$ by assumption, there is a preference $R'_i \in {\cal R}_i$ such that $f(\hat{R}_{-i}, R'_i) = z$. Therefore, $\{x,z\} \subseteq O_j((\hat R_{N \setminus \{i,j\}}, R'_i),{\cal R}_j)$. This implies by Lemma \ref{free} that $\{x, z\} \in S^{-1}({\cal R}_j)$.
Second, setting $S = \{j\}$, and taking into account that $\bar{R}_k = \hat{R}_k$ for all $k \not\in D_{r(f)}(\bar{R}_{-i}, \hat{R}_{-i})$ by condition (d), we can see that $O_i((\hat{R}_j, \bar{R}_{N \setminus \{i, j\}}), {\cal R}_i) = x$. Thus, for all $R_i \in {\cal R}_i$, $f(\hat{R}_j, \bar{R}_{N \setminus \{i, j\}}, R_i) = x$. Since $O_i(\bar{R}_{-i}, {\cal R}_{-i}) = \{x, y\}$ by assumption, there is a preference $R''_i \in {\cal R}_i$ such that $f(\bar{R}_{-i}, R''_i) = y$. Therefore, $\{x,y\} \subseteq O_j((\bar R_{N \setminus \{i,j\}}, R''_i), {\cal R}_j)$. Hence, by Lemma \ref{free}, $\{x, y\} \in S^{-1}({\cal R}_j)$.
\end{proof}

\noindent Consider for each agent $j \in D_{r(f)}(\bar{R}_{-i}, \hat{R}_{-i})$ two preferences $R_j^4$ and $R_j^5$ such that $z \, P_j^4 \, y \, P_j^4 \, x$, $y \, P_j^5 \, z \, P_j^5 \, x$ and $v \, P_l \, w$ for all $(v, w) \in S({\cal R}_j) \setminus \{(y, z), (z, y)\}$ and all $l \in \{4, 5\}$. 
By Lemma \ref{libres}, either $R_j^4$ or $R_j^5$ (or both) belong to ${\cal R}_j$. 
Divide the agents of $D_{r(f)}(\bar{R}_{-i}, \hat{R}_{-i})$ into the following four groups: 
$$S_1 = \{j \in D_{r(f)}(\bar{R}_{-i}, \hat{R}_{-i}) : y \, \bar{P}_j \, z \mbox{ and } y \, \hat{P}_j \, z\}.$$
$$S_2 = \{j \in D_{r(f)}(\bar{R}_{-i}, \hat{R}_{-i}) : y \, \bar{P}_j \, z \mbox{ and } z \, \hat{P}_j \, y\}.$$
$$S_3 = \{j \in D_{r(f)}(\bar{R}_{-i}, \hat{R}_{-i}) : z \, \bar{P}_j \, y \mbox{ and } y \, \hat{P}_j \, z\}.$$
$$S_4 = \{j \in D_{r(f)}(\bar{R}_{-i}, \hat{R}_{-i}) : z \, \bar{P}_j \, y \mbox{ and } z \, \hat{P}_j \, y\}.$$

\noindent Note that if $j \in S_1 \cup S_2 \cup S_3$, then $R_j^5 \in {\cal R}_j$.
Similarly, if $j \in S_2 \cup S_3 \cup S_4$, then $R_j^4 \in {\cal R}_j$.

\medskip

\noindent First, construct the following sequence of subprofiles.
Starting at $\bar{R}_{-i}$, change the preferences of all agents $j \in S_1 \cup S_2$ one-by-one from $\bar{R}_j$ to $R_j^5$; after that, change the preferences of all agents $j \in S_3 \cup S_4$ one-by-one from $\bar{R}_j$ to $R_j^4$. We show that in all subprofiles of this sequence, the option set for agent $i$ in the preference domain ${\cal R}_i$ has to be $\{x, y\}$. 
Suppose by contradiction that this is not the case. 
Consider the subprofile of the sequence in which, for the last time in the sequence, the option set for agent $i$ in the preference domain ${\cal R}_i$ is $\{x, y\}$ and the next subprofile of the sequence. That is, consider a subprofile $\tilde{R}_{-i} \in {\cal D}_{-i}$ of the sequence such that $O_i(\tilde{R}_{-i}, {\cal R}_i) = \{x, y\}$ and an agent $j \in D_{r(f)}(\bar{R}_{-i}, \hat{R}_{-i})$ with a preference $\tilde{R}'_j \in \{R_j^4, R_j^5\}$ such that $O_i((\tilde{R}_{-\{i, j\}}, \tilde{R}'_j), {\cal R}_i) \neq \{x, y\}$. Since $O_i(\tilde{R}_{-i}, {\cal R}_i) = \{x, y\}$, $O_i(\tilde{R}_{-i}, {\cal R}'_i) = x$ by Lemma \ref{range0}. Then, by Lemma \ref{xesta}, $x \in O_i((\tilde{R}_{-\{i, j\}}, \tilde{R}'_j), {\cal R}'_i)$. Thus, $x \in O_i((\tilde{R}_{-\{i, j\}}, \tilde{R}'_j), {\cal R}_i)$ by Lemma \ref{range0}. Since $O_i((\tilde{R}_{-\{i, j\}}, \tilde{R}'_j), {\cal R}_i) \neq \{x,y\}$ by assumption, we can conclude that $O_i((\tilde{R}_{-\{i, j\}}, \tilde{R}'_j), {\cal R}_i) \in \{ \{x\} ,\{x,z\} \{x,y,z\} \}$. 
First, suppose that $O_i((\tilde{R}_{-\{i, j\}}, \tilde{R}'_j), {\cal R}_i) = \{x,y,z\}$.
It follows from Lemma \ref{range0} that $O_i((\tilde{R}_{-\{i, j\}}, \tilde{R}'_j), {\cal R}'_i) = \{x, z\}$. This contradicts Lemma \ref{proposition}.
Second, suppose that $O_i((\tilde{R}_{-\{i, j\}}, \tilde{R}'_j), {\cal R}_i) = \{x,z\}$.
Then, by Lemma \ref{range0}, $O_i((\tilde{R}_{-\{i, j\}}, \tilde{R}'_j), {\cal R}'_i) = \{x,z\}$.
Since we have already seen that $O_i(\tilde{R}_{-i}, {\cal R}_i)=\{x,y\}$ and $O_i(\tilde{R}_{-i}, {\cal R}'_i)=\{x\}$, we can apply Lemma \ref{pre-proposition} to obtain that $O_i((\tilde{R}_{-\{i, j\}}, \tilde{R}'_j), {\cal R}_i) = \{x,y,z\}$.
This contradicts Lemma \ref{proposition}.
Third, suppose that $O_i((\tilde{R}_{-\{i, j\}},$ $\tilde{R}'_j), {\cal R}_i) = x$.
Then, for all $R_i \in {\cal R}_i, f(R_i, \tilde{R}_{-\{i, j\}}, \tilde{R}'_j) = x$.
Moreover, it follows from $O_i(\tilde{R}_{-i}, {\cal R}_i) = \{x, y\}$ that there is a preference $R'_i \in {\cal R}_i$ such that $f(R'_i, \tilde R_{-i})=y$. Agent $j$ thus manipulates $f$ at $(R'_i,  \tilde{R}_{-\{i, j\}}, \tilde{R}'_j)$ via $\tilde{R}_j$. This contradicts that $f$ is SP. Hence, the option set for agent $i$ in the preference domain ${\cal R}_i$ for all subprofiles along the sequence is $\{x, y\}$.

\medskip

\noindent Second, construct the following different sequence of subprofiles.
Starting at $\hat{R}_{-i}$, change the preferences of all agents $j \in S_1 \cup S_3$ one-by-one from $\hat{R}_j$ to $R_j^5$; after that, change the preferences of all agents $j \in S_2 \cup S_4$ one-by-one from $\hat{R}_j$ to $R_j^4$. 
We show that in all subprofiles of this sequence, the option set for agent $i$ in the preference domain ${\cal R}_i$ has to include $z$ but not $y$. 
Suppose first by contradiction that $z$ is not in the option set for agent $i$ in some subprofile of the sequence. 
Consider the subprofile of the sequence in which, for the last time in the sequence, the option set for agent $i$ in the domain ${\cal R}_i$ includes $z$ and the next subprofile of the sequence. 
That is, consider a subprofile $\tilde{R}_{-i} \in {\cal D}_{-i}$ such that $z \in O_i(\tilde{R}_{-i}, {\cal R}_i)$ and an agent $j \in D_{r(f)}(\bar{R}_{-i}, \hat{R}_{-i})$ with a preference $\tilde{R}'_j \in \{R_j^4, R_j^5\}$ such that $O_i((\tilde{R}_{-\{i, j\}}, \tilde{R}'_j), {\cal R}_i) \subseteq \{x, y\}$. Then, by Lemma \ref{range0}, $O_i((\tilde{R}_{-\{i, j\}}, \tilde{R}'_j), {\cal R}'_i) = x$ and $z \in O_i(\tilde{R}_{-i}, {\cal R}'_i)$. Therefore, for all $R_i \in {\cal R}'_i$, $f(R_i, \tilde{R}'_j, \tilde{R}_{-\{i, j\}}) = x$. 
Also, there is a preference $R'_i \in {\cal R}'_i$ such that $f(R'_i, \tilde{R}_{-i}) = z$. 
Agent $j$ thus manipulates $f$ at $(R'_i, \tilde{R}'_j, \tilde{R}_{-\{i, j\}})$ via $\tilde{R}_j$. 
This contradicts that $f$ is SP and, thus, we can conclude that in all subprofiles of this sequence, the option set for agent $i$ in the preference domain ${\cal R}_i$ has to include $z$. 
Suppose next by contradiction that $y$ belongs to the option set for agent $i$ in the preference domain ${\cal R}_i$ for some subprofile of this sequence. That is, consider a subprofile $\tilde{R}_{-i} \in {\cal D}_{-i}$ such that $y \not \in O_i(\tilde{R}_{-i}, {\cal R}_i)$ and an agent $j \in D_{r(f)}(\bar{R}_{-i}, \hat{R}_{-i})$ with a preference $\tilde{R}'_j \in \{R_j^4, R_j^5\}$ such that $y \in O_i((\tilde{R}_{-\{i, j\}}, \tilde{R}'_j), {\cal R}_i)$. Then, by Lemma \ref{range0}, $x \in O_i((\tilde{R}_{-\{i, j\}}, \tilde{R}'_j), {\cal R}_i)$. Since we have already seen that $z$ belongs to the option set of agent $i$ when her preference domain is ${\cal R}_i$ along the whole sequence of subprofiles, $O_i((\tilde{R}_{-\{i, j\}}, \tilde{R}'_j), {\cal R}_i) =\{x,y,z\}$.
Then, by Lemma \ref{range0}, $O_i((\tilde{R}_{-\{i, j\}}, \tilde{R}'_j), {\cal R}'_i) =\{x,z\}$, contradicting Lemma \ref{proposition}. We can thus conclude that in all subprofiles of this sequence, the option set for agent $i$ in the preference domain ${\cal R}_i$ is $\{z\}$ or $\{x, z\}$.
Then, by Lemma \ref{range0}, this option set coincides with the option set for agent $i$ in the preference domain ${\cal R}'_i$.

\medskip

\noindent In order to simplify notation, denote from now on $S_i \cup S_j \equiv S_{ij}$ and $D_{r(f)}(\bar R_{-i},\hat R_{-i}) \equiv S$.
Observe that the first sequence ends at the subprofile $(\bar{R}_{-S}, R_{S_{12}}^5, R_{S_{34}}^4)$, while the second sequence ends at the subprofile $(\hat{R}_{-S}, R_{S_{13}}^5, R_{S_{24}}^4)$. 
Thus, we have already established that
\begin{itemize}
\item[] $O_i(\bar{R}_{-S}, R_{S_{12}}^5, R_{S_{34}}^4), {\cal R}'_i) = x$, $O_i((\bar{R}_{-S}, R_{S_{12}}^5, R_{S_{34}}^4), {\cal R}_i) = \{x, y\}$ and
\item []$O_i((\hat{R}_{-S}, R_{S_{13}}^5, R_{S_{24}}^4), {\cal R}'_i)$ $= O_i((\hat{R}_{-S}, R_{S_{13}}^5, R_{S_{24}}^4), {\cal R}_i) \in \{\{z\}, \{x, z\}\}$. 
\end{itemize} 
It can be also checked that $d_{r(f)}((\bar{R}_{-S}, R_{S_{12}}^5, R_{S_{34}}^4),$ $(\hat{R}_{-S}, R_{S_{13}}^5, R_{S_{24}}^4)) \leq d_{r(f)}(\bar{R}_{-i}, \hat{R}_{-i})$.
%, with strict inequality if $S_1 \cup S_4 \neq \emptyset$}. 

\medskip

\noindent If $S_2 = S_3 = \emptyset$, then $(\bar{R}_{-S}, R_{S_{12}}^5, R_{S_{34}}^4)=(\hat{R}_{-S}, R_{S_{13}}^5, R_{S_{24}}^4)$.
This contradicts that $O_i((\bar{R}_{-S},$ $R_{S_{12}}^5, R_{S_{34}}^4), {\cal R}_i) \neq O_i((\hat{R}_{-S}, R_{S_{13}}^5, R_{S_{24}}^4), {\cal R}_i)$. Thus, $S_2 \cup S_3 \neq \emptyset$.

\medskip

\noindent Suppose first that $|S_2 \cup S_3| = 1$. Suppose that $S_2 = \{j\}$ and $S_3 = \emptyset$ (the proof is similar if $S_2=\emptyset$ and $S_3=\{j\}$). Observe that the subprofiles $(\bar{R}_{-S}, R_{S_{12}}^5, R_{S_{34}}^4)$ and $(\hat{R}_{-S}, R_{S_{13}}^5, R_{S_{24}}^4)$ only differ because of agent $j$ and, thus, $(\bar{R}_{-S}, R_{S_{12}}^5, R_{S_{34}}^4) = (\hat{R}_{-S}, R_{S_{13} \cup \{j\}}^5, R_{S_{24} \setminus \{j\}}^4)$. Given that $z \in O_i((\hat{R}_{-S}, R_{S_{13}}^5, R_{S_{24}}^4), {\cal R}'_i)$, there is a preference $R_i \in {\cal R}'_i$ such that $f(R_i, \hat{R}_{-S},$ $R_{S_{13}}^5, R_{S_{24}}^4) = z$. 
We can then use the fact that $O_i((\bar{R}_{-S}, R_{S_{12}}^5, R_{S_{34}}^4), {\cal R}'_i) = x$ in order to see that $f(R_i, \bar{R}_{-S}, R_{S_{12}}^5, R_{S_{34}}^4) = f(R_i, \hat{R}_{-S}, R_{S_{13} \cup \{j\}}^5, R_{S_{24} \setminus \{j\}}^4) = x$. Then, agent $j$ manipulates $f$ at $(R_i, \hat{R}_{-S}, R_{S_{13} \cup \{j\}}^5, R_{S_{24} \setminus \{j\}}^4)$ via $R_j^4$.

\medskip

\noindent Suppose now that $|S_2 \cup S_3| > 1$. Consider any agent $j \in S_2$ (the proof is similar if $S_2 = \emptyset$ and $j \in S_3$) and the subprofile $(\bar{R}_{-S}, R_{S_{12} \setminus \{j\}}^5, R_{S_{34} \cup \{j\}}^4)$. Since $r(f)=\{x,y,z\}$, it follows from Lemma \ref{range0} that $O_i(\bar{R}_{-S}, R_{S_{12} \setminus \{j\}}^5, R_{S_{34} \cup \{j\}}^4), {\cal R}_i) \in \{ \{x\}, \{z\}, \{x,z\}, \{x,y\}, \{x,y,z\} \}$.

\begin{itemize}
\item Suppose that $O_i((\bar{R}_{-S}, R_{S_{12} \setminus \{j\}}^5, R_{S_{34} \cup \{j\}}^4), {\cal R}_i) \in \{\{z\},\{x,z\}\}$.
Then, by Lemma \ref{range0}, $O_i((\bar{R}_{-S}, R_{S_{12} \setminus \{j\}}^5, R_{S_{34} \cup \{j\}}^4), {\cal R}'_i) = O_i((\bar{R}_{-S}, R_{S_{12} \setminus \{j\}}^5, R_{S_{34} \cup \{j\}}^4), {\cal R}_i)$. 
We can observe that $d_{r(f)}((\bar{R}_{-S},R_{S_{12}}^5, R_{S_{34}}^4), (\bar{R}_{-S}, R_{S_{12} \setminus \{j\}}^5, R_{S_{34} \cup \{j\}}^4)) = 1$ and that this value is strictly smaller than $d_{r(f)}((\bar{R}_{-S}, R_{S_{12}}^5,$ $R_{S_{34}}^4), (\hat{R}_{-S}, R_{S_{13}}^5, R_{S_{24}}^4))= |S_2 \cup S_3|$. Thus, $d_{r(f)}((\bar{R}_{-S},R_{S_{12}}^5, R_{S_{34}}^4), (\bar{R}_{-S}, R_{S_{12} \setminus \{j\}}^5, R_{S_{34} \cup \{j\}}^4)) < d_{r(f)}(\bar{R}_{-i}, \hat{R}_{-i})$ and $d_{r(f)}(\bar{R}_{-i}, \hat{R}_{-i})$ is not minimal, which contradicts condition (c).

\item Suppose that $O_i((\bar{R}_{-S}, R_{S_{12} \setminus \{j\}}^5, R_{S_{34} \cup \{j\}}^4), {\cal R}_i) =\{x,y\}$.
It follows from Lemma \ref{range0} that $O_i((\bar{R}_{-S}, R_{S_{12} \setminus \{j\}}^5, R_{S_{34} \cup \{j\}}^4), {\cal R}'_i) = x$. 
At this point we can observe that $d_{r(f)}((\bar{R}_{-S}, R_{S_{12} \setminus \{j\}}^5, R_{S_{34} \cup \{j\}}^4), ( \hat{R}_{-S}, R_{S_{13}}^5, R_{S_{24}}^4)) = d_{r(f)}((\bar{R}_{-S}, R_{S_{12}}^5, R_{S_{34}}^4), (\hat{R}_{-S}, R_{S_{13}}^5,$ $R_{S_{24}}^4)) - 1$. Then, $d_{r(f)}((\bar{R}_{-S}, R_{S_{12} \setminus \{j\}}^5, R_{S_{34} \cup \{j\}}^4), (\hat{R}_{-S}, R_{S_{13}}^5, R_{S_{24}}^4)) < d_{r(f)}(\bar{R}_{-i}, \hat{R}_{-i})$. Thus, $d_{r(f)}(\bar{R}_{-i}, \hat{R}_{-i})$ is not minimal, which contradicts condition (c).

\item Suppose that $O_i((\bar{R}_{-S}, R_{S_{12} \setminus \{j\}}^5, R_{S_{34} \cup \{j\}}^4), {\cal R}_i) =\{x,y,z\}$.
It follows then from Lemma \ref{range0} that $O_i((\bar{R}_{-S}, R_{S_{12} \setminus \{j\}}^5, R_{S_{34} \cup \{j\}}^4), {\cal R}_i) =\{x,z\}$.
This contradicts Lemma \ref{proposition}.

\item Suppose that $O_i((\bar{R}_{-S}, R_{S_{12} \setminus \{j\}}^5, R_{S_{34} \cup \{j\}}^4), {\cal R}_i) =\{x\}$. 
We can thus conclude that for all $R_i \in {\cal R}_i$, $f(R_i, \bar{R}_{-S}, R_{S_{12} \setminus \{j\}}^5, R_{S_{34} \cup \{j\}}^4)=x$. Since $O_i((\bar{R}_{-S},R_{S_{12}}^5, R_{S_{34}}^4), {\cal R}_i) = \{x, y\}$, there is a preference $R'_i \in {\cal R}_i$ such that $f(R'_i, \bar{R}_{-S},R_{S_{12}}^5, R_{S_{34}}^4) = y$. Agent $j$ then manipulates $f$ at $(R'_i, \bar{R}_{-S}, R_{S_{12} \setminus \{j\}}^5, R_{S_{34} \cup \{j\}}^4)$ via $R_j^5$. This contradicts that $f$ is SP.
\end{itemize}

\noindent This concludes the proof of PART 3 and of the theorem.

\section*{Calculations of Example 1}

\noindent Since $|X|=5$, there are $5! = 120$ rankings in ${\cal R}$. 
The 60 rankings in which $y \, P \, x$ are all included in ${\cal R}_i$, yet if $x \, P \, y$, then only those rankings in which $z$ is the top alternative of the triple $\{z, w, v\}$ belong to ${\cal R}_i$ (1/3 of 60 rankings). Since ${\cal R}_i$ contains 80 rankings and $|N|=2$,  there are $80^2 = 6400$ possible preference profiles and $5^{6400}$ social choice rules. 

\medskip

\noindent To calculate how many of the $5^{6400}$ rules comply with the two-step procedure, observe that the first step asks the two agents about her preference between $x$ and $y$. So, there are four possible response profiles in this first step. We next determine how many of the possible subrules are in each of the cases dictatorial or strategy-proof of range 2.

\medskip

\noindent {\bf Both agents report that they prefer $y$ to $x$: 59 subrules}

\begin{itemize}
\item {\it 5 dictatorial subrules of range 1.} This is because $|X|=5$.

\item {\it 36 strategy-proof subrules of range 2.} Since $|X|=5$, there are 10 pairs of alternatives. The range cannot be $\{x, y\}$ because preferences on that pair have been determined in the first step. For each of the 9 feasible pairs, there are 4 possible subrules: 2 dictatorial subrules and the 2 subrules in which $s \in \{x,y\}$ is selected unless both agents agree that $t \in \{x,y\} \setminus s $ is better than $s$. 

\item {\it 14 dictatorial subrules of range 3.} There are 2 possible dictators and 10 possible triples of alternatives.  The range cannot be $\{x, y, z\}$, $\{x, y, w\}$, or $\{x, y, v\}$ because the dictator does not choose $x$ in the presence of $y$.

\item {\it 4 dictatorial subrules of range 4.} There are 2 possible dictators and 5 possible quadruples of alternatives. The range cannot be  $\{x, y, z, v\}$, $\{x, y, z, w\}$, or $\{x, y, w, v\}$ because the dictator does not choose $x$ in the presence of $y$.

\item {\it 0 dictatorial subrules of range 5.} This is because the dictator does not choose $x$.

\end{itemize}

\noindent {\bf Both agents report that they prefer $x$ to $y$: 37 subrules}

\medskip

\begin{itemize}
\item {\it 5 dictatorial subrules of range 1.} This is because $|X|=5$.

\item {\it 28 strategy-proof subrules of range 2.} The range cannot be $\{x, y\}$, $\{z, w\}$, or $\{z, v\}$ because the preferences on these pairs have been determined in the first step (directly for $\{x, y\}$ and indirectly otherwise). For each of the 7 feasible pairs, there are 4 possible subrules: 2 dictatorial subrules and the 2 subrules in which $s \in \{x,y\}$ is selected unless both agents agree that $t \in \{x,y\} \setminus s$ is better than $s$. 

\item {\it 4 dictatorial subrules of range 3.} There are 2 possible dictators and 10 possible triples of alternatives. Since the dictator does not choose $y$ in the presence of $x$ nor $w$ or $v$ in the presence of $z$, the range can only be $\{x, w, v\}$ or $\{y, w, v\}$. 

\item {\it 0 dictatorial subrules of range 4 or 5.} This is because the dictator does not choose $y$ in the presence of $x$ nor $w$ or $v$ in the presence of $z$.
\end{itemize}

\medskip

\noindent {\bf The reports of the two agents are different: 46 subrules}

\medskip

\noindent Suppose that agent 1 is the one who reports that she prefers $x$ to $y$.

\begin{itemize}
\item {\it 5 dictatorial subrules of range 1.} This is because $|X|=5$.

\item {\it 30 strategy-proof subrules of range 2.} Since $|X|=5$, there are 10 pairs of alternatives. The range cannot be $\{x, y\}$ because preferences on that pair have been determined in the first step. If the range is $\{z, w\}$ or $\{z, v\}$, agent 2 is necessarily a dictator because agent 1 has already indirectly revealed in the first step that she prefers $z$ to both $w$ and $v$. For each of the remaining 7 feasible pairs, there are 4 possible subrules: 2 dictatorial subrules and the 2 subrules in which $s \in \{x,y\}$ is selected unless both agents agree that $t \in \{x,y\} \setminus s$ is better than $s$. 

\item {\it 9 dictatorial subrules of range 3.}  There are 2 possible dictators and 10 possible triples of alternatives. If agent 1 is the dictator, $y$ is not chosen in the presence of $x$ nor is $w$ or $v$ chosen in the presence of $z$. Therefore, the range can only be $\{x, w, v\}$ or $\{y, w, v\}$. If agent 2 is the dictator, $x$ is not chosen in the presence of $y$. Hence, the range cannot be $\{x, y, z\}$, $\{x, y, v\}$, or $\{x, y, w\}$.

\item  {\it 2 dictatorial subrules of range 4.} There are 2 possible dictators and 5 possible quadruples of alternatives. If agent 1 is the dictator, $y$ is not chosen in the presence of $x$ nor is $w$ or $v$ chosen in the presence of $z$. This eliminates all quadruples. If agent 2 is the dictator, $x$ is not chosen in the presence of $y$. Hence, the range can be $\{x, z, w, v\}$ or $\{y, z, w, v\}$.

\item {\it 0 dictatorial subrules of range 5.}  This is because $x$ and $y$ cannot be simultaneously in the range.
\end{itemize}

We conclude that $59 \cdot 37 \cdot 46^2$ rules are consistent with the two-step procedure.

\section*{Calculations of Example 2}

\noindent Since $|X|=5$, there are $5! = 120$ rankings in ${\cal R}$. 
There is 1 single-peaked ranking with peak at $v$ and another one with peak at $z$, there are 4 single-peaked rankings with peak at $w$ and peak at $y$ each, and there are 6 single-peaked rankings with peak at $x$. Thus, ${\cal R}_j$ includes 16 rankings. Given that $|N|=2$, there are $16^2 = 256$ preference profiles and $5^{256}$ social choice rules.

\medskip

\noindent To calculate how many of the $5^{256}$ rules comply with the two-step procedure, observe that the first step asks the two agents about their peaks. The 4 possible answers are: at $v$, at $w$, at $x$, or to the right of $x$. So, there are 16 possible response profiles in this first step. Note that in each of the 16 cases, there are 5 dictatorial subrules of range 1 and that there are no dictatorial subrules that have at least range 3. The latter is due to the fact that there is no triple of alternatives for which an agent has three rankings in which each of the alternatives is the best alternative of the triple. To complete the calculations, we determine how many subrules are in each case strategy-proof of range 2.

\medskip

\noindent {\bf The agents report the same peak}.

\begin{itemize}
\item {\it 0 subrules for peaks at $v$.} This is because the preferences of both agents are entirely determined in the first step.

\item {\it 12 subrules for peaks at $w$.} The range can only be $\{v, x\}$, $\{v, y\}$, or $\{v, z\}$. For each of the 3 feasible pairs, there are 4 possible subrules: 2 dictatorial subrules and the 2 subrules in which $s \in \{x,y\}$ is selected unless both agents agree that $t \in \{x,y\} \setminus s$ is better than $s$. 

\item {\it 16 subrules for peaks at $x$.} The range can only be $\{v, y\}$, $\{v, z\}$, $\{w, y\}$, or $\{w, z\}$. For each of the 4 feasible pairs, there are 4 possible subrules: 2 dictatorial subrules and the 2 subrules in which $s \in \{x,y\}$ is selected unless both agents agree that $t \in \{x,y\} \setminus s$ is better than $s$. 

\item {\it 16 subrules for peaks to the right of $x$.} The range can only be $\{v, z\}$, $\{w, z\}$, $\{x, z\}$, or $\{y, z\}$. For each of the 4 feasible pairs, there are 4 possible subrules: 2 dictatorial subrules and the 2 subrules in which $s \in \{x,y\}$ is selected unless both agents agree that $t \in \{x,y\} \setminus s$ is better than $s$. 
\end{itemize}

\noindent {\bf The agents report different peaks}\medskip

Suppose that the reported peak of agent 1 is to the left of that of agent 2.

\begin{itemize}
\item {\it 3 subrules for peaks at $v$ and $w$.} The range can only be $\{v, x\}$, $\{v, y\}$, or  $\{v, z\}$. The only possible subrule for each range is that agent 2 is the dictator.

\item {\it 4 subrules for peaks at $v$ and $x$.} The range can only be $\{v, y\}$, $\{v, z\}$, $\{w, y\}$, or $\{w, z\}$. The only possible subrule for each range is that agent 2 is the dictator.

\item {\it 4 subrules for peaks at $v$ and to the right of $x$.} The range can only be $\{v, z\}$, $\{w, z\}$, $\{x, z\}$, or $\{y, z\}$. The only possible subrule for each range is that agent 2 is the dictator.

\item {\it 11 subrules for peaks at $w$ and $x$.} The range cannot be $\{v, w\}$, $\{w, x\}$, $\{x, y\}$, $\{x, z\}$, or $\{y, z\}$. If the range is $\{v, x\}$, then agent 1 is a dictator. If the range is $\{w, y\}$ or $\{w, z\}$, then agent 2 is a dictator. Finally, if the range is $\{v, y\}$ or $\{v, z\}$, there are 4 possible subrules: 2 dictatorial subrules and the 2 subrules in which $s \in \{x,y\}$ is selected unless both agents agree that $t \in \{x,y\} \setminus s$ is better than $s$. 

\item {\it 9 subrules for peaks at $w$ and to the right of $x$.} The range cannot be $\{v, w\}$, $\{w, x\}$, $\{w, y\}$, or $\{x, y\}$. If the range is $\{v, x\}$ or $\{v, y\}$, then agent 1 is a dictator. If the range is $\{w, z\}$, $\{x, z\}$, or $\{y, z\}$, then agent 2 is a dictator. If the range is $\{v, z\}$, there are 4 possible subrules: 2 dictatorial subrules and the 2 subrules in which $s \in \{x,y\}$ is selected unless both agents agree that $t \in \{x,y\} \setminus s$ is better than $s$. 

\item {\it 12 subrules for peaks at $x$ and to the right of $x$.}  The range cannot be $\{v, w\}$, $\{v, x\}$, $\{w, x\}$, or $\{x, y\}$. If the range is $\{v, y\}$ or $\{w,y\}$, then agent 1 is a dictator. If the range is $\{x, z\}$ or $\{y, z\}$, then agent 2 is a dictator. If the range is $\{v, z\}$ or $\{w, z\}$, there are 4 possible subrules: 2 dictatorial subrules and the 2 subrules in which $s \in \{x,y\}$ is selected unless both agents agree that $t \in \{x,y\} \setminus s$ is better than $s$. 
\end{itemize}

\noindent We conclude that $(5+0) \cdot (5+3)^2 \cdot (5+4)^4 \cdot (5+9)^2 \cdot (5+11)^2 \cdot (5+12)^3 \cdot (5+16)^2$ rules are consistent with the two-step procedure.

\end{document}